\documentclass[pra,twocolumn,showpacs,floatfix]{revtex4-1}

\usepackage[utf8]{inputenc}
\usepackage{graphicx}
\usepackage{flafter}

\usepackage{amsmath,amssymb,amstext}
\usepackage{braket}

\usepackage{float}
\usepackage{color}
\usepackage{subfigure}

\newcommand{\ii}{\mathrm{i}}

\let \Re \relax
\DeclareMathOperator{\Re}{Re}
\let \Im \relax
\DeclareMathOperator{\Im}{Im}

\newcommand{\bk}[1]{\left(#1\right)}					
\newcommand{\Bk}[1]{\left[#1\right]}

\newcommand{\absq}[1]{\left|#1\right|^2}				
		
\newcommand{\euler}[1]{\text{e}^{#1}}					
			
\definecolor{checkedGray}{gray}{.4}

\begin{document}
\title{Symmetry-breaking oscillations in membrane optomechanics}
\author{C. Wurl}
\affiliation{Institut f{\"u}r Physik,
Ernst-Moritz-Arndt-Universit{\"a}t Greifswald, 17487 Greifswald, Germany }

\author{A. Alvermann}
\thanks{Corresponding author}
\email{alvermann@physik.uni-greifswald.de}
\affiliation{Institut f{\"u}r Physik,
Ernst-Moritz-Arndt-Universit{\"a}t Greifswald, 17487 Greifswald, Germany }

\author{H. Fehske}
\affiliation{Institut f{\"u}r Physik,
Ernst-Moritz-Arndt-Universit{\"a}t Greifswald, 17487 Greifswald, Germany }

\begin{abstract}
We study the classical dynamics of a membrane inside a cavity in the situation where this optomechanical system possesses a reflection symmetry.
Symmetry breaking occurs through supercritical and subcritical pitchfork bifurcations of the static fixed point solutions.
Both bifurcations can be observed through variation of the laser-cavity detuning,
which gives rise to a boomerang-like fixed point pattern with hysteresis.
The symmetry-breaking fixed points evolve into self-sustained oscillations when the laser intensity is increased.
In addition to the analysis of the accompanying Hopf bifurcations we describe these oscillations at finite amplitudes with an ansatz that fully accounts for the frequency shift relative to the natural membrane frequency.
We complete our study by following the route to chaos for the membrane dynamics.
\end{abstract}

\pacs{}

\maketitle

\section{Introduction}

\begin{figure}
\hspace*{-0.5cm}
\includegraphics[width=0.9\linewidth]{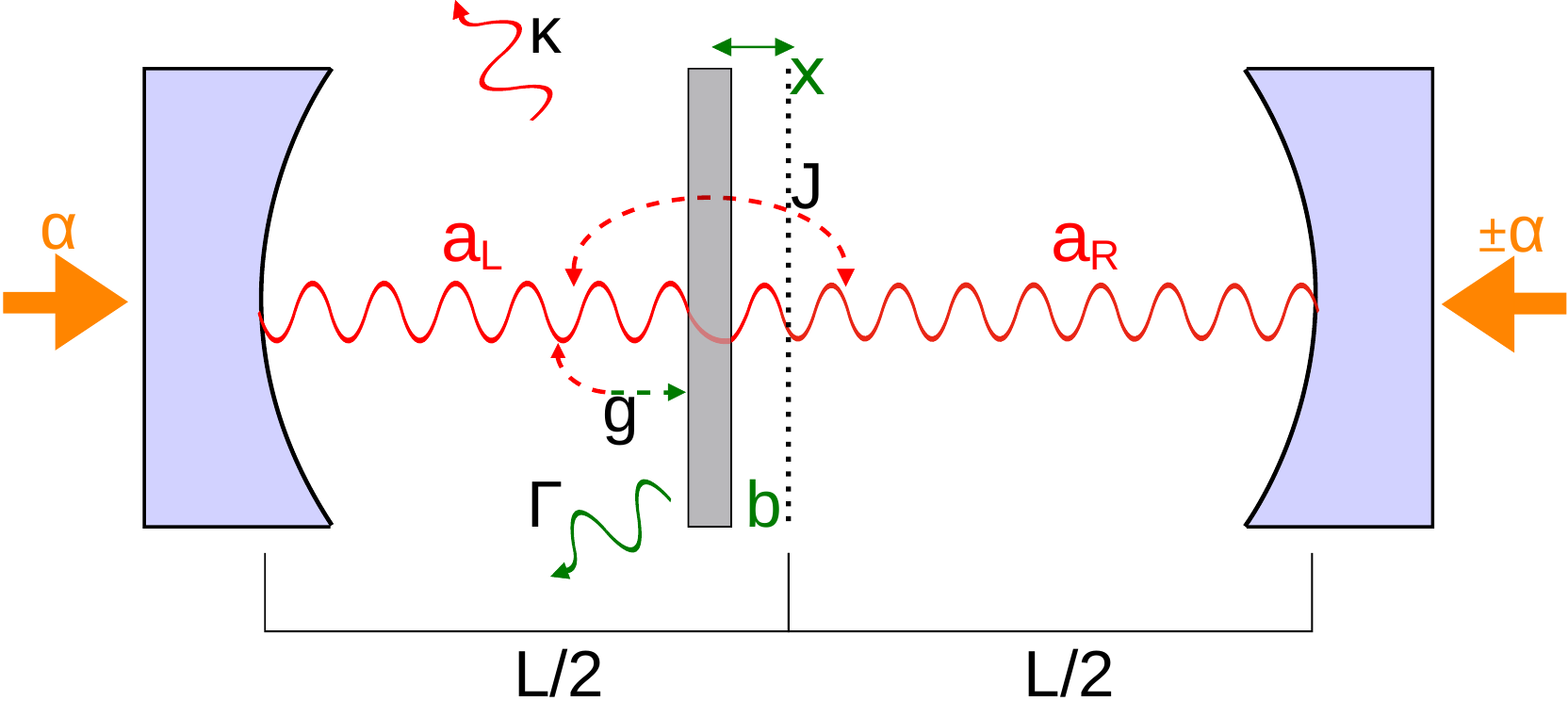}
\caption{(Color online) The ``membrane-in-the-middle'' setup consists of a vibrating, partially reflective membrane placed in the center of a cavity, which is pumped by an external laser through the left and right mirrors with equal intensities.
}
\label{fig:Setup}
\end{figure}

Optomechanical systems~\cite{KV08,MG09,M13,AKM13} show a variety of dynamical patterns in the classical and quantum regime~\cite{HH10,WH13}.
Several aspects of the classical nonlinear dynamics of these systems
have been studied theoretically and observed in the experiment,
including self-sustained oscillations~\cite{RKCV05,KRCSV05,CRYKV05,MHG06}, multistability and hysteresis~\cite{LKM08}, and chaotic~\cite{CCV07,BAF14_PRL,LJMW15} behaviour.

A different line of inquiry concerns the modification of the classical dynamics due to quantum effects~\cite{Gutz90}.
The general correspondence between the classical and quantum dynamics of optomechanical systems, 
and the specific fate of self-sustained oscillations under the influence of quantum noise~\cite{GZ04} and phase space diffusion~\cite{Schl01} have been addressed in recent studies~\cite{ALPVBK16,WLG16,SA16}. 
These studies require a clear picture of the classical dynamical patterns,
to be able to identify the influence of quantum effects.

In order to contribute to this picture we address in this paper the nonlinear dynamics of a membrane inside a cavity (see Fig.~\ref{fig:Setup}),
with a focus on the self-sustained oscillations that break the reflection symmetry of the specific setup considered here.
Our work is motivated by previous studies of similar setups that addressed, e.g., the symmetry breaking at zero detuning~\cite{MO15}, the onset of chaotic motion~\cite{LH11}, or pattern formation and buckling phase transitions for a flexible membrane~\cite{XKFRLT15,RN16}.
We extend these studies along three lines.
First, we analyse the pitchfork and saddle-node bifurcations related to symmetry breaking and hysteresis,
which leads to a clear characterization of the different transitions between the symmetric and non-symmetric situation. 
Second, we establish a scaling relation for the bifurcations and fixed-point solutions that allows for tuning the symmetry-breaking transitions to different parameter regimes.
Third, we introduce a new ansatz for the self-sustained membrane oscillations, and develop an intuitive physical picture of symmetry-breaking oscillations that is based on the power balance between optical and mechanical degrees of freedom associated with this ansatz.
These results should help to observe static and dynamical symmetry breaking in future experiments. 
The close relation between our theoretical findings and the actual experiment is established by the translation rules between model and real `physical' parameters given below.

One specific result of our study with potential experimental relevance is that the frequency of the self-sustained oscillations is shifted significantly relative to the natural membrane frequency.
This is in contrast to the ``cantilever-cavity'' system with one photon mode, where self-sustained oscillations occur approximately at the cantilever frequency~\cite{MHG06}. The frequency shift, which 
can be determined experimentally from the position of the optical sidebands, contains additional information about system parameters such as the membrane stiffness. At least in principle, mechanical parameters could thus be obtained from optical frequency measurements.

\section{Theoretical setup}

The symmetric ``membrane-in-the-middle'' setup considered here consists of a membrane with high reflectivity placed near the cavity center (see Fig.~\ref{fig:Setup}).
Two degenerate photon modes in the left and right half of the cavity contribute equally to the radiation pressure acting on the membrane.
Photon tunneling through the membrane connects both photon modes, lifts their degeneracy,
and results in a quadratic dispersion of the optical modes as a function of the membrane position~\cite{TZ07,JS08,BU08,LS12}.

For the theoretical analysis of this situation it is convenient to work with dimensionless quantities
(see App.~\ref{app:EOM} for a summary), especially,
to measure time in units of the inverse membrane frequency ($\Omega^{-1}$).
Then, the classical equations of motion read 
\begin{subequations}%
\label{EOM}
\begin{align}
\dot{x} & = \phantom{-} p  \;,  \label{EOM1} \\
\dot{p} & =  - x  -  \Gamma p -  g\bk{\absq{ a_{\text{L}}}-\absq{a_{\text{R}}}} \;, \label{EOM2} \\
\dot{a}_{\text{L}} & =  \Bk{\ii  \Delta - \ii x -  \kappa}  a_{\text{L}}- \ii  J a_{\text{R}} - \ii \;, \label{EOM3} \\
\dot{ a}_{\text{R}} & = \Bk{\ii \Delta + \ii x - \kappa} a_{\text{R}}- \ii  J  a_{\text{L}} -  \ii \label{EOM4} \;,
\end{align}
\end{subequations}
for the membrane position ($x$) and momentum ($p$) and the photon field amplitudes in the left ($a_L$) and right ($a_R$) cavity.
These equations contain five dimensionless parameters:
the laser-cavity detuning $\Delta=\bk{\Omega_{\text{las}}-\Omega_{\text{cav}}}/\Omega$, 
cavity decay rate $\kappa=\pi c / (2 F L \Omega)$, 
mechanical damping $\Gamma=1/Q_{\text{m}}$,
membrane transmissivity $J=\euler{i\varphi} \sqrt{2\bk{1-r}} \, (c/L) /\Omega$,
and effective radiation pressure $g=(\pi c \, \Omega_\text{cav} P)/(m \Omega^5 L^3 F)$.
These parameters are obtained from the `physical' parameters of the cavity (length $L$,
frequency $\Omega_\mathrm{cav}$, finesse $F$), the membrane (frequency $\Omega$,
mass $m$, quality factor $Q_{\text{m}}$, reflectivity $r$), and the laser (frequency $\Omega_\mathrm{las}$, transmitted power $P$, phase difference $\varphi$) as specified here and in App.~\ref{app:EOM}.

Note that the above equations of motion are valid for a relative phase $e^{\ii \varphi} = \pm 1$ of the laser amplitude at the right and left mirror, with $ J > 0$ ($ J < 0$) for equal (opposite) phase.
The laser power enters through the parameter $g$.
For typical experimental setups from the literature~\cite{AKM13}, we have $g \lesssim 10$ with significant optical losses ($\kappa \simeq 1$) and small mechanical damping ($\Gamma \simeq 10^{-4} \ll 1$). 
Since the effective optomechanical coupling $g$ can be adjusted via the laser power, different experimental implementations are conceivable to achieve sufficiently large values of $g$. In the optomechanical setup in Ref.~\cite{TZ07}, for example, a pump power on the order of $P\sim 10^{-8}\text{W}$ is required if the cavity is driven with laser light with frequency $\Omega_{\text{las}}/2\pi\sim 10^{14}\text{Hz}$. However, possible experimental realizations depend on the availability of highly reflective membranes with very small $J$.

We now study the fixed point bifurcations related to symmetry breaking (Sec.~\ref{sec:SymmBreak}),
the Hopf bifurcations leading to self-sustained oscillations and the properties of these oscillations at finite amplitudes (Sec.~\ref{sec:Osci}),
before we follow the route to chaos (Sec.~\ref{sec:Chaos}) and conclude immediately thereafter (Sec.~\ref{sec:Conc}). The appendices collect additional information on the derivation of the dimensionless equations of motion (App.~\ref{app:EOM}), the stability analysis (App.~\ref{app:Jac}),
and the finite amplitude ansatz (App.~\ref{app:Fourier}).

\section{Symmetry breaking}
\label{sec:SymmBreak}

The equations of motion~\eqref{EOM} are invariant under the replacement $x \mapsto -x$ (with $p \mapsto -p$ and swapping $a_{L/R} \mapsto a_{R/L}$),
which defines the reflection symmetry of the system with respect to the membrane position.
The symmetry implies the existence of a trivial fixed point $x_0 = 0$,
while symmetry breaking results in additional nontrivial fixed points $\pm x_i \ne 0$.

The fixed points are obtained from Eq.~\eqref{EOM} as the solutions
with $\dot{x}=\dot{p}=\dot{a}_{\text{L}}=\dot{a}_{\text{R}}=0$.
Four nontrivial fixed points can exist in addition to $x_0=0$,
namely,
\begin{subequations}
\label{FixedPoints}
\begin{align}
 x_{1/2} &=\pm \sqrt{-\gamma+2\sqrt{f}} \;, \\
 x_{3/4} &=\pm \sqrt{-\gamma-2\sqrt{f}} \;,
\end{align}
\end{subequations}
where $\gamma= \kappa^{2}+ J^{2} -  \Delta^{2}$ and $f=-  \Delta^{2} \kappa^{2}-  g \bk{  \Delta+ J}$. 
These fixed points exist if the respective terms under the square root are non-negative.
As a consequence of the reflection symmetry they occur in pairs $\pm x_i$ with opposite sign.
The corresponding values for $a_{\mathrm L/R}$ are
\begin{equation}\label{FixedPointsValueA}
 a_{\mathrm L/\mathrm R} = \frac{\Delta \pm  x + \ii  \kappa +  J}{(\ii \Delta -  \kappa)^2 +  x^2 +  J^2} 
\end{equation}
for all fixed points, with the $+$ ($-$) sign for $a_L$ ($a_R$).

\paragraph*{Pitchfork bifurcation}
As $g$ is increased, the nontrivial fixed points appear through a pitchfork bifurcation at 
\begin{equation}\label{GPitchfork}
  g_p = -\frac{\Delta^2 \kappa^2}{\Delta +  J} - \frac{1}{4} \frac{(\kappa^2 +  J^2 - \Delta^2)^2}{\Delta +  J} \;.
\end{equation}

\begin{figure}
\hspace*{\fill}
\includegraphics[width=0.46\linewidth]{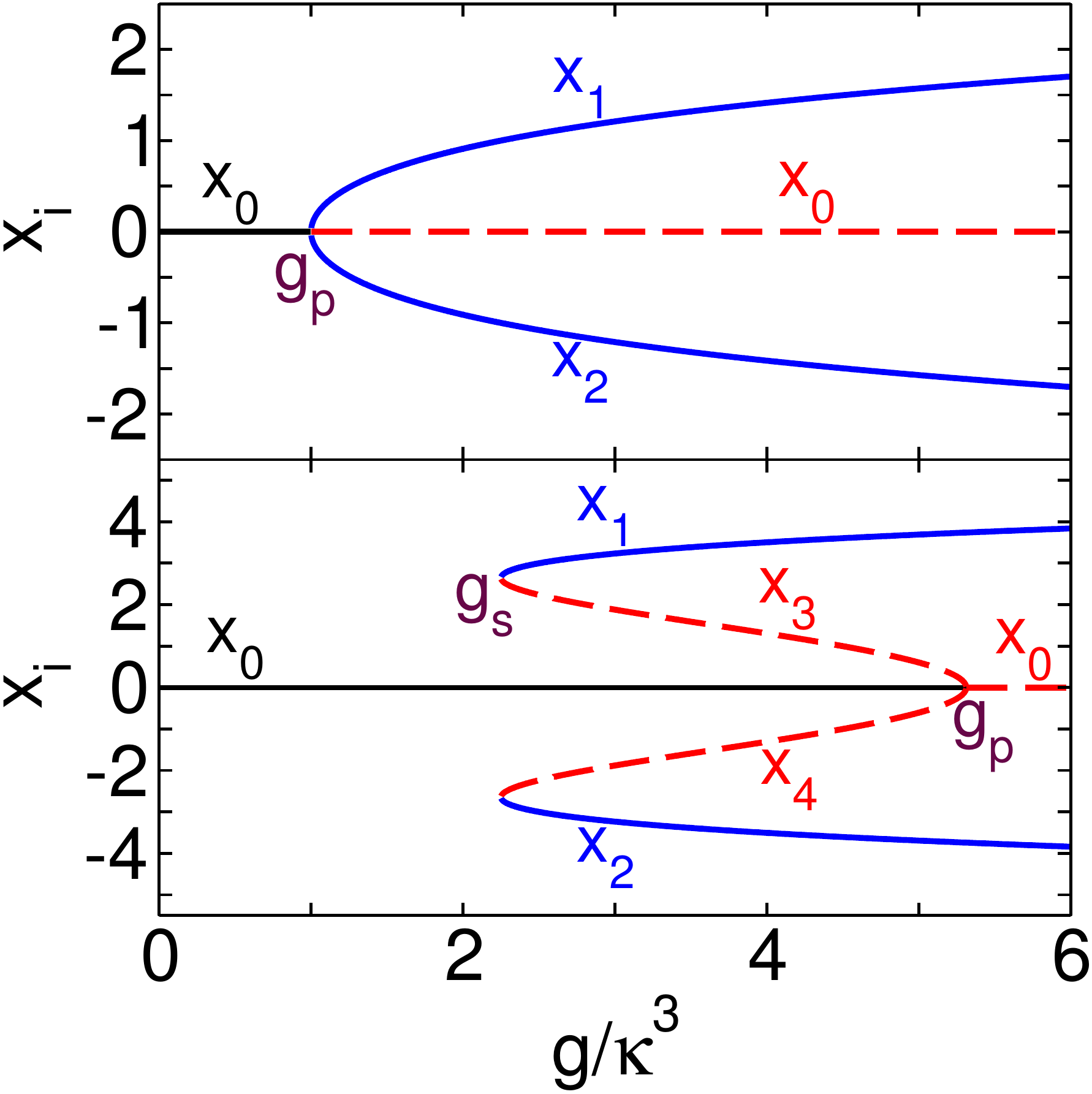}
\hspace*{\fill}
\includegraphics[width=0.47\linewidth]{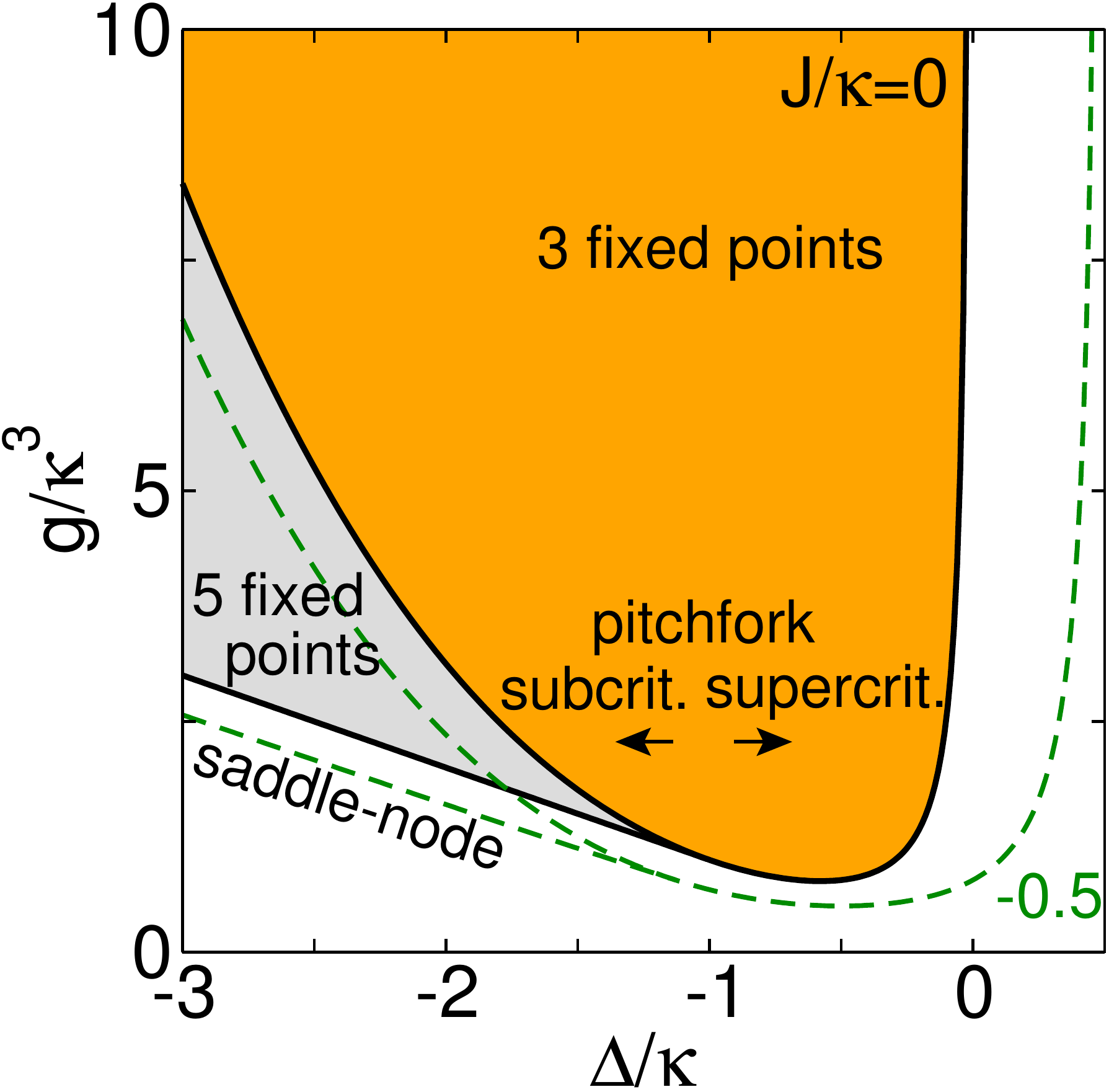}
\hspace*{\fill}
\caption{(Color online) Left panel: Supercritical pitchfork bifurcation for small detuning (upper plot, for $\Delta/\kappa=0$), and subcritical pitchfork and saddle-node bifurcation for large detuning (lower plot, for $\Delta/\kappa =-3$), both for $J/\kappa=-1$.
Right panel: Diagram of bifurcations (at $g_p$ and $g_s$) and number of fixed points
in the $g$-$\Delta$ plane, for $J/\kappa=0$.
Small non-zero $J/\kappa$ shifts, essentially, the boundary curves in the plane
(see, e.g., the dashed curves for $J/\kappa = -0.5$).
Nontrivial fixed points exist for $\Delta < - J$.
}
\label{fig:bifurcation}
\end{figure}

For small detuning $|\Delta| \le \sqrt{\kappa^2 +  J^2}$
the bifurcation at $g_p$ is a supercritical pitchfork bifurcation
(see Fig.~\ref{fig:bifurcation}, left panel, upper plot).
For $ g < g_p$ only the trivial fixed point $x_0 = 0$ exists.
For  $ g >  g_p$, the trivial fixed point becomes unstable and the two stable fixed points $x_1$, $x_2$ appear.
For large detuning $|\Delta| \ge \sqrt{\kappa^2 +  J^2}$
the bifurcation at $g_p$ is a subcritical pitchfork bifurcation (see Fig.~\ref{fig:bifurcation}, left panel, lower plot),
where the two unstable fixed points $x_3$, $x_4$ exist together with the stable trivial fixed point $x_0$ for $g< g_p$.
The pitchfork bifurcation is accompanied by a saddle-node bifurcation at
\begin{equation}\label{GAllFive}
g_s = - \frac{ \Delta^2 \kappa^2}{\Delta +  J} \;,
\end{equation}
which connects the unstable fixed points $x_3$, $x_4$ to the two stable fixed points $x_1$, $x_2$.
For $g_s < g < g_p$ all five fixed points coexist.

\paragraph*{Scaling}

Eqs.~\eqref{FixedPoints}--\eqref{GAllFive} are invariant under the scaling $\Delta \mapsto s \Delta$, $J \mapsto s J$, $\kappa \mapsto s \kappa$, $x \mapsto s x$,   $a_{L/R} \mapsto (1/s) a_{L/R}$, $g_i \mapsto s^3 g_i$, for any $s>0$.
Therefore, the positions of the fixed points depend only on the appropriate ratios, e.g., $J/\kappa$, $\Delta/\kappa$, and $g/\kappa^3$.
The stability of the fixed points, however, depends on the absolute values of the system parameters and changes with $s$ (see Sec.~\ref{sec:Osci}).
The occurrence of nontrivial fixed points is summarized in Fig.~\ref{fig:bifurcation}.

Note that for $\Delta=0$ only the supercritical pitchfork bifurcation occurs.
In this situation symmetry breaking is formally related to the superradiant phase transition in the Dicke model~\cite{MO15}. 

\paragraph*{Boomerang pattern}

\begin{figure}
\hspace*{\fill}
\includegraphics[width=0.47\linewidth]{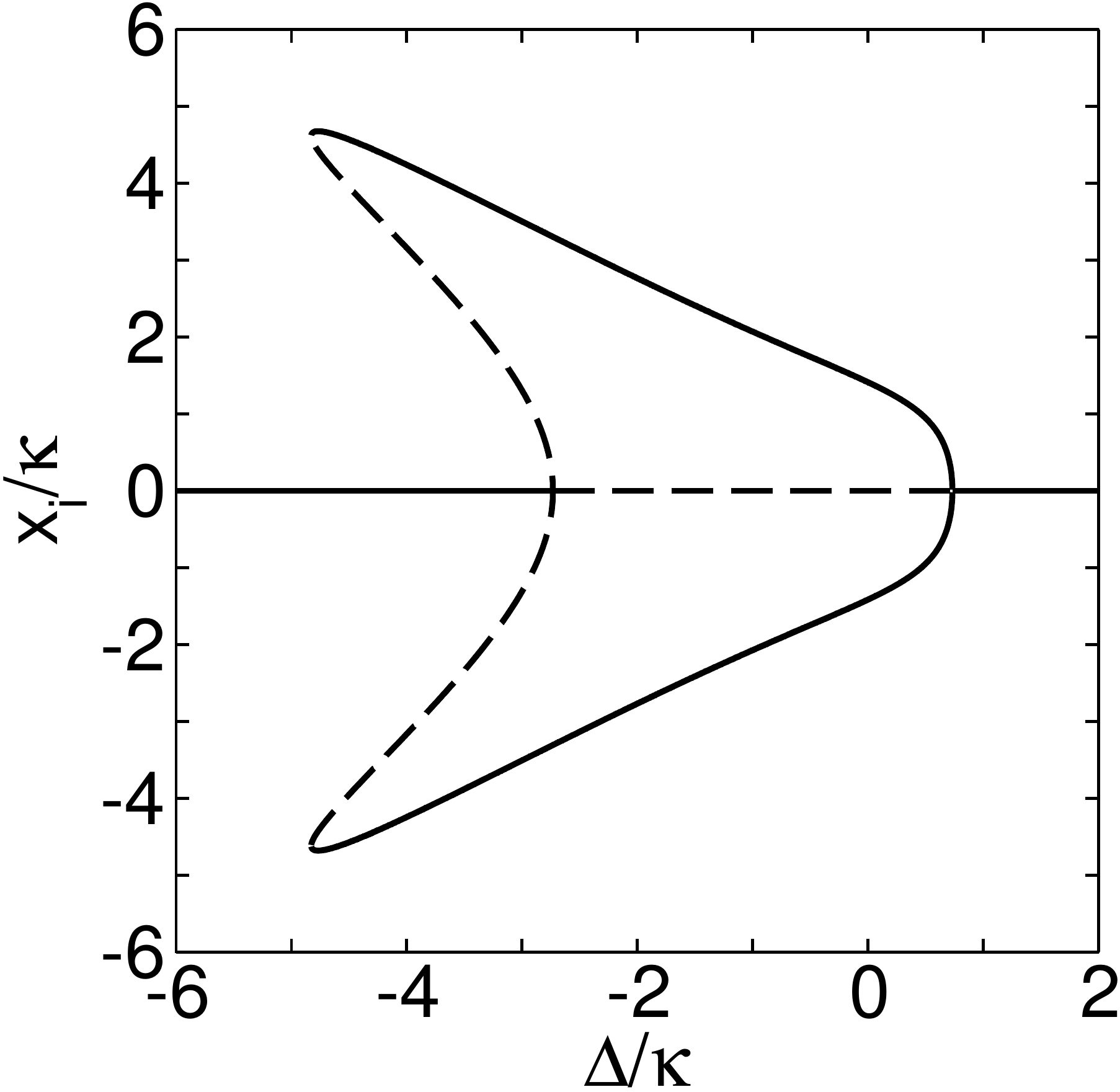}
\hspace*{\fill}
\includegraphics[width=0.46\linewidth]{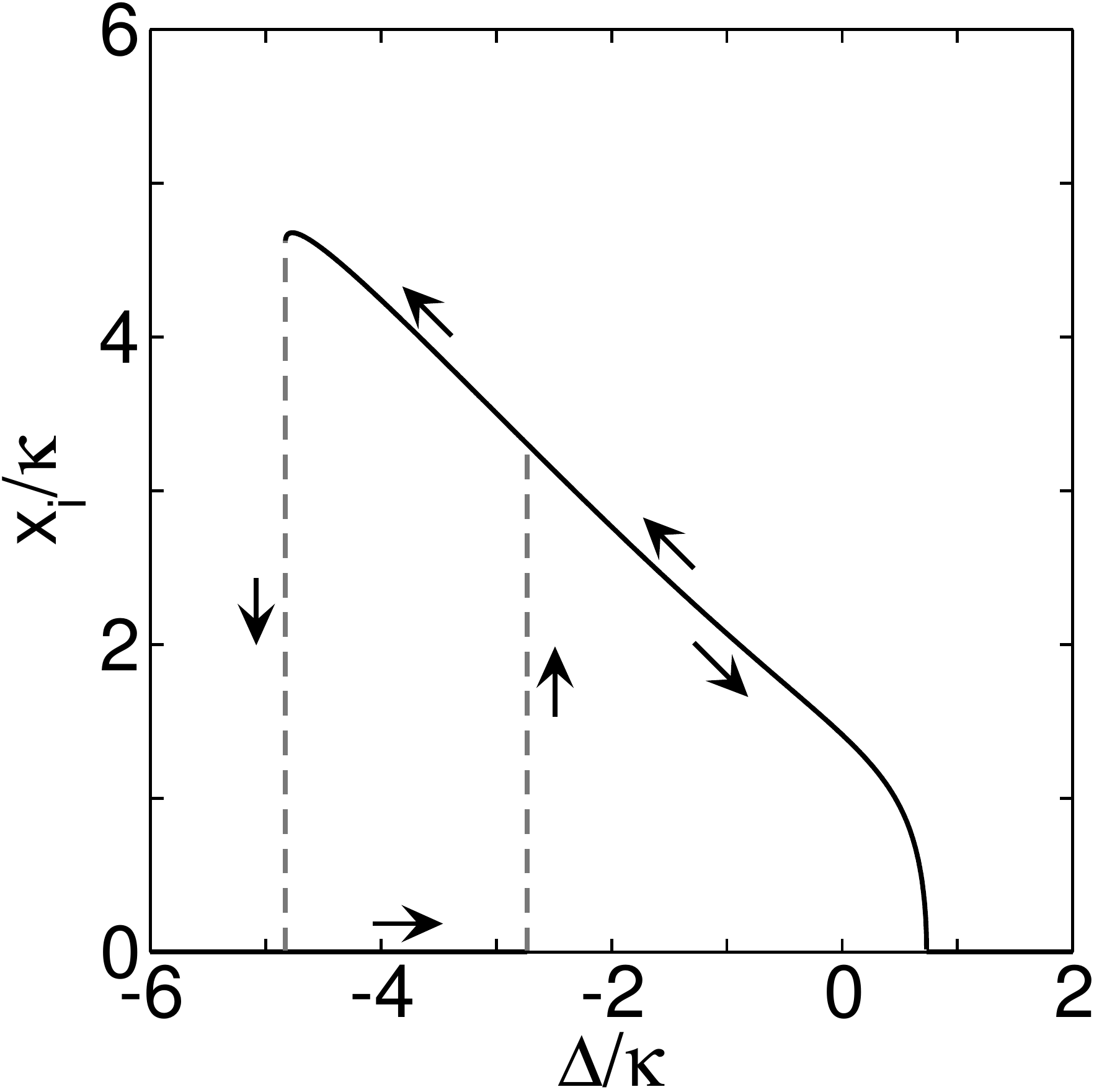}
\hspace*{\fill}
\caption{Left panel: ``Boomerang'' fixed point pattern as a function of $\Delta$,
for fixed $g/\kappa^3 = 4$, $J/\kappa = -1$.
Right panel: Hysteresis of fixed points for cyclic change of $\Delta/\kappa$.
The stable (solid curves) and unstable (dashed curves) fixed points as drawn here are obtained for small $g$ (i.e., small scaling parameter $s$). At larger $g$, fixed points can lose stability through Hopf bifurcations (cf. Sec.~\ref{sec:Osci}).
}
\label{fig:boomerang}
\end{figure}

Changing the laser-cavity detuning $\Delta$ instead of the effective radiation pressure $g$ 
allows for observation of the supercritical and subcritical pitchfork bifurcation in succession (see Fig.~\ref{fig:boomerang}, left panel).
The saddle-node bifurcation in the resulting ``boomerang''-like fixed point pattern can be observed through the hysteresis that occurs when $\Delta$ is changed along a cycle (see Fig.~\ref{fig:boomerang}, right panel).

\section{Self-sustained oscillations}
\label{sec:Osci}

\subsection{Hopf bifurcations}

In the vicinity of the pitchfork and saddle-node bifurcations the stability of fixed points changes according to the type of the bifurcation.
Away from the fixed point bifurcations additional dynamical Hopf bifurcations can occur,
through which potentially stable fixed points are replaced by oscillatory orbits.

\begin{figure}
\center
\includegraphics[scale=0.198]{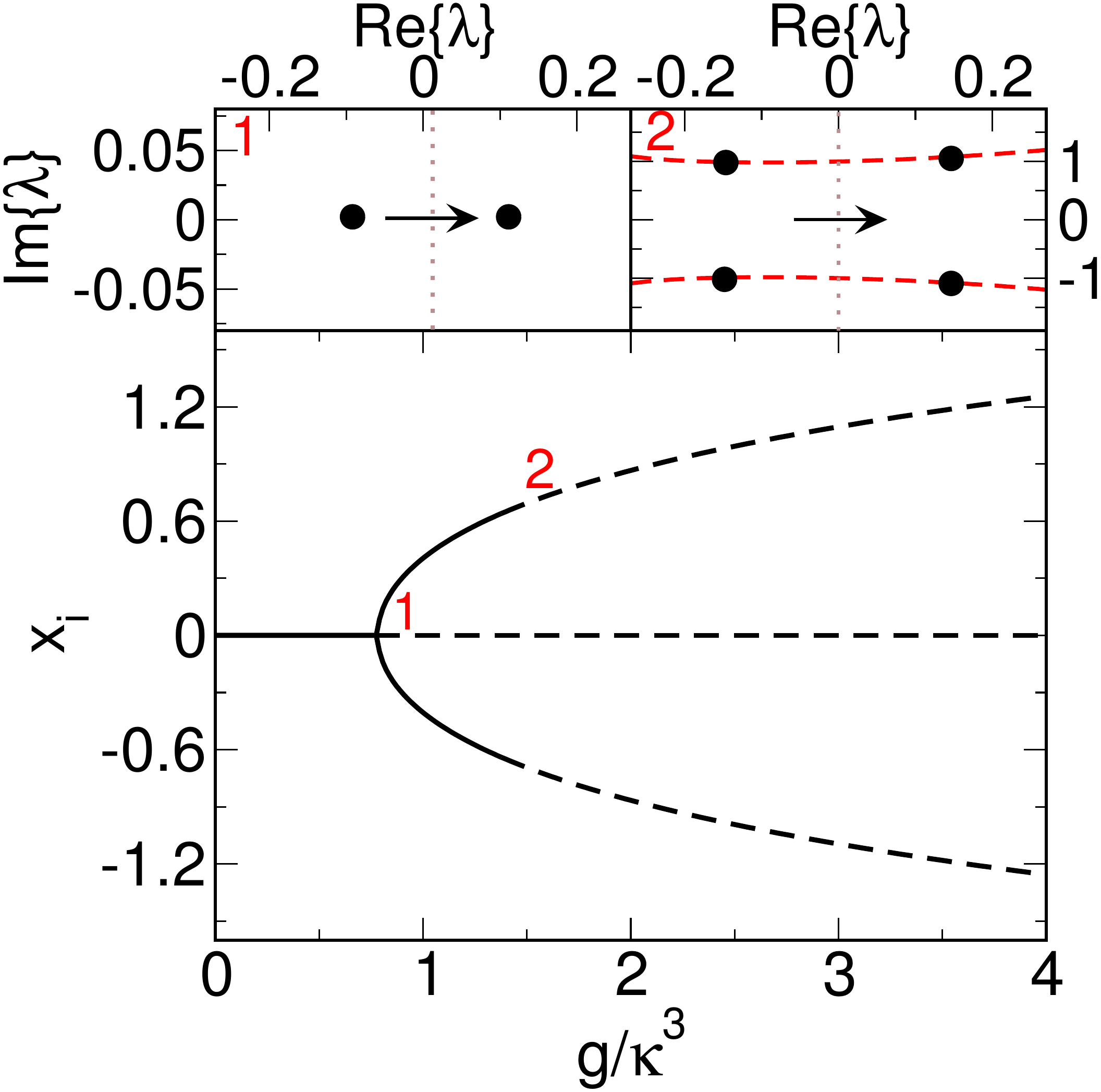}
\includegraphics[scale=0.198]{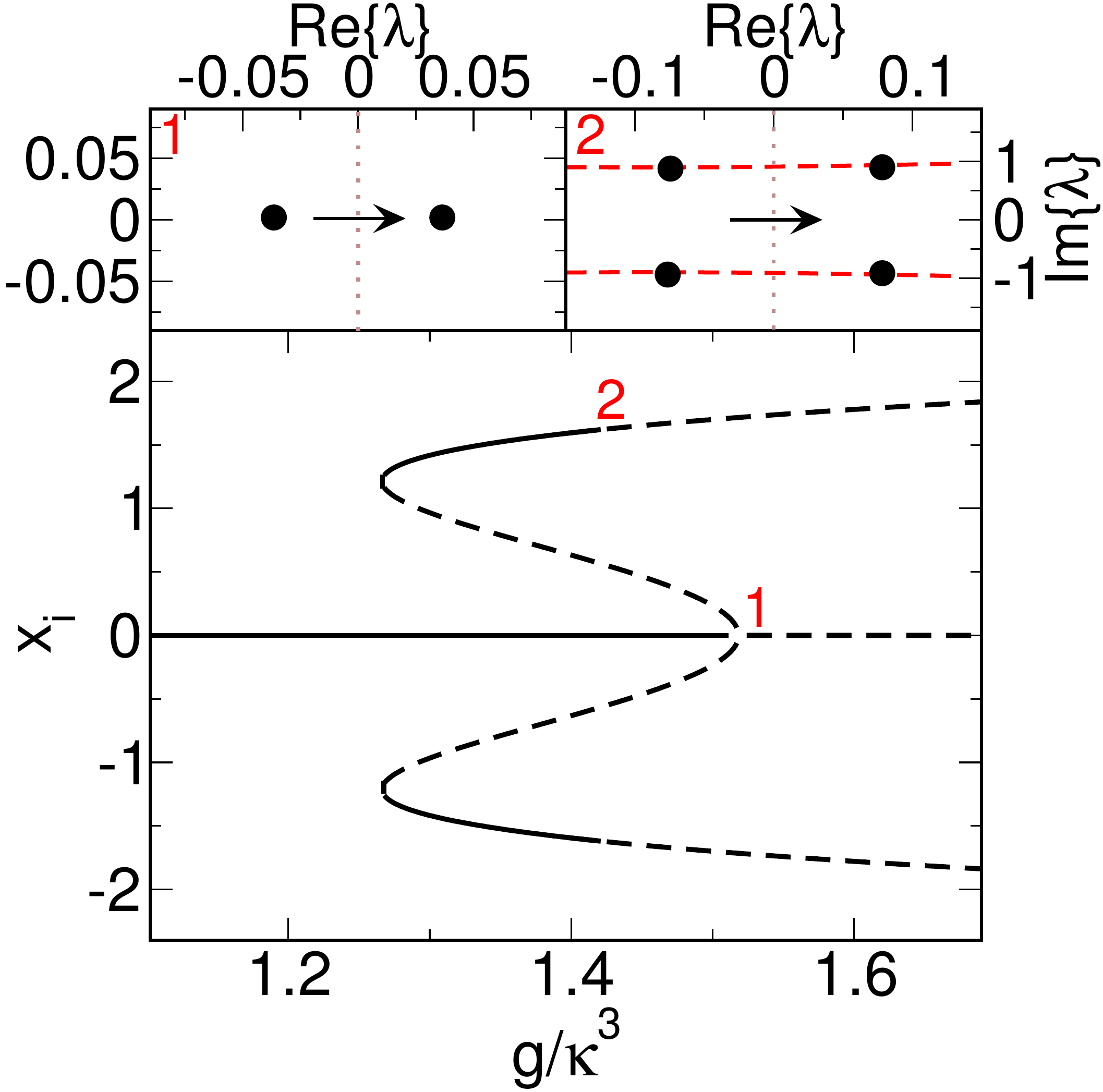}
\caption{(Color online)  Stability characteristics for the supercritical 
(left panel, $\Delta/\kappa=0$) and subcritical 
(right panel, $\Delta/\kappa=-1.65$) pitchfork bifurcation, for $J/\kappa=-0.5$ and $\kappa=1$. Stable (unstable) fixed points are plotted as solid (dashed) curves. The red numbers indicate the pitchfork (1) and Hopf bifurcations (2), which can be distinguished by the number of eigenvalues of the stability matrix that cross the imaginary axis (upper panels).
\label{fig:Pitchfork}}
\end{figure}

The stability of the fixed points is determined by the stability matrix that is obtained from linearization of the equations of motion (see App.~\ref{app:Jac} for explicit expressions).
Fig.~\ref{fig:Pitchfork} shows the stability of fixed points according to the linear analysis for the supercritical and subcritical pitchfork bifurcation.
Note that the stability changes under the $s$-scaling that leaves the fixed points invariant,
such that we have to specify the absolute value of, e.g., $\kappa$ in Fig.~\ref{fig:Pitchfork}.

\begin{figure}
\hspace*{\fill}
\includegraphics[width=0.47\linewidth]{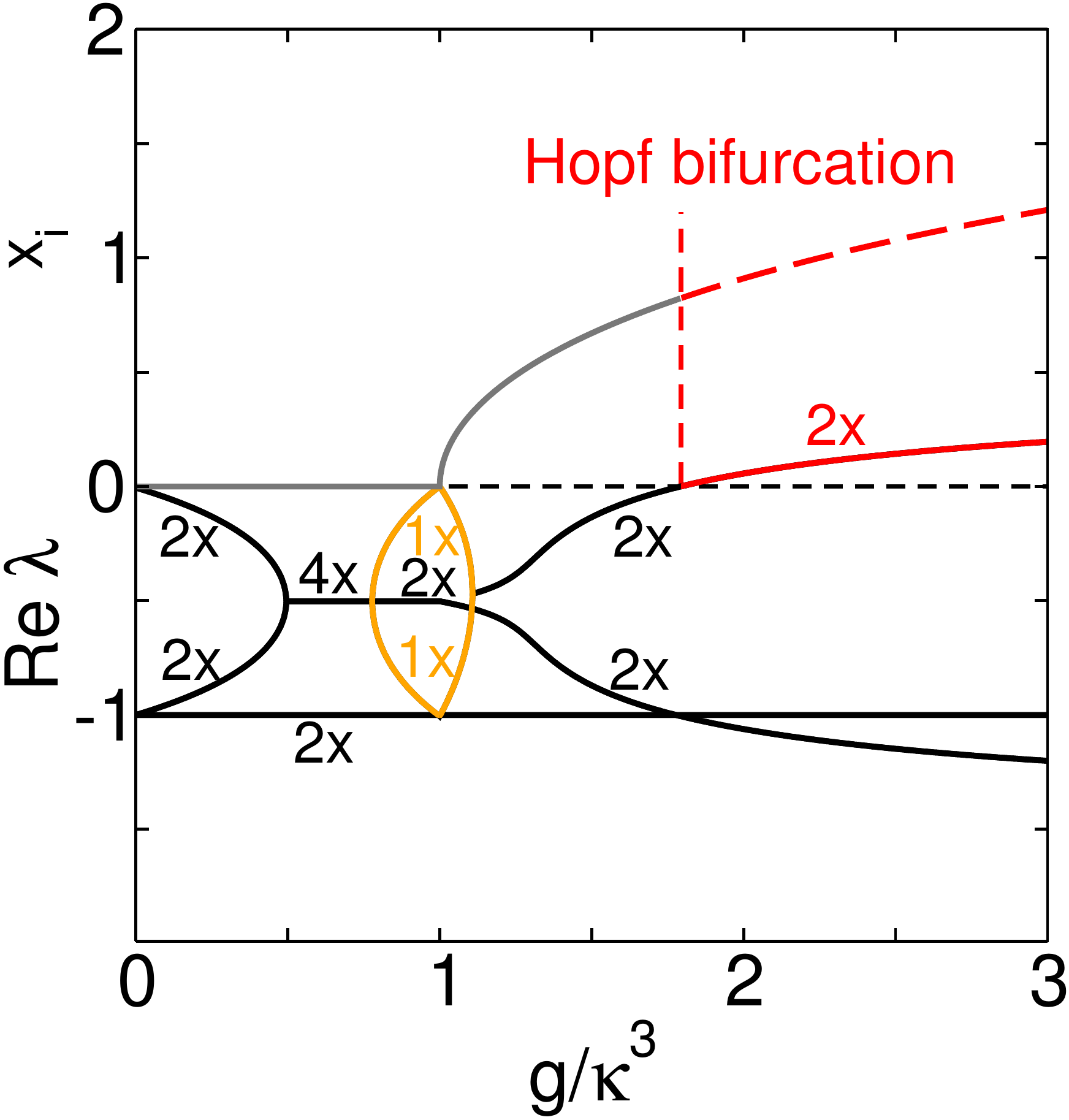}
\hspace*{\fill}
\includegraphics[width=0.47\linewidth]{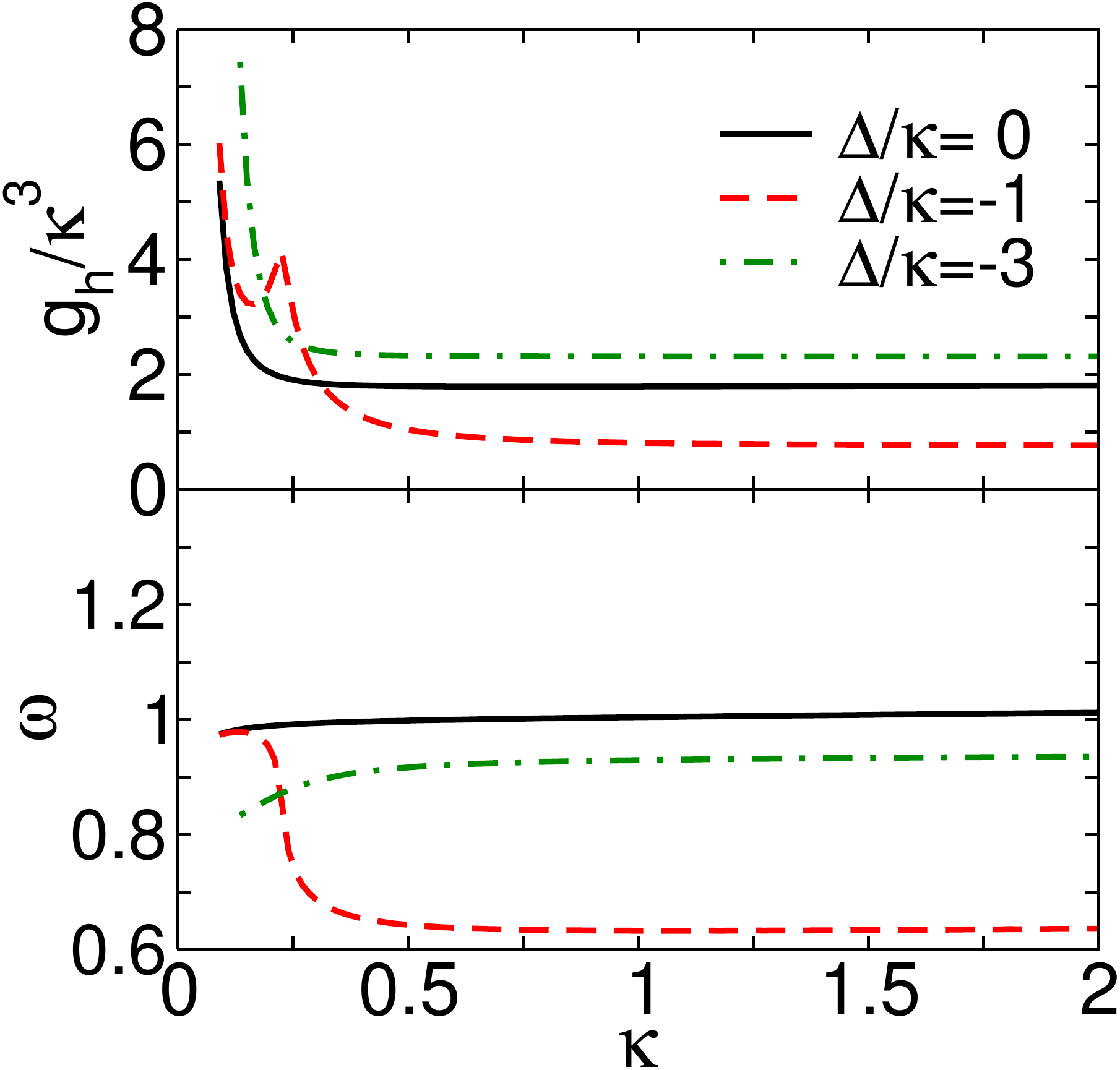}
\hspace*{\fill}
\caption{(Color online) Left panel: Real part of the six eigenvalues of the stability matrix across the pitchfork bifurcation, for $\Delta/\kappa=0$, $J/\kappa=-1$ as in Fig.~\ref{fig:bifurcation}, and $\kappa = 1$. 
Small numbers indicate the multiplicity of the eigenvalues. At $g \approx 1.79$ the fixed point $x_1$ loses stability through a Hopf bifurcation.
Right panel: Position of the Hopf bifurcation ($g_h$) and frequency of the small amplitude oscillations ($\omega$) as a function of the absolute value of $\kappa$,
for $\Delta/\kappa = 0,-1,-3$.
}
\label{fig:realparts}
\end{figure}

Fig.~\ref{fig:realparts} shows the real part of the eigenvalues of the stability matrix, following the fixed points $x_0 \to x_1$ through the supercritical pitchfork bifurcation at small $|\Delta|$.
In the vicinity of $g_p$ we observe how one real eigenvalue touches the imaginary axis 
($\Re \lambda = 0$)
at the bifurcation.
At a certain value $g > g_p$
a pair of complex conjugate eigenvalues ($\lambda$, $\lambda^*$) crosses the imaginary axis,
and a (supercritical) Hopf bifurcation takes place.
The frequency of the oscillations that appear immediately after the Hopf bifurcation is given by the imaginary parts of the eigenvalue pair.

The position of the Hopf bifurcation and the oscillation frequency depend on the absolute parameter values,
not only the ratios $J/\kappa$ etc., 
and thus change under the $s$-scaling that leaves the fixed point pattern invariant.
Fig.~\ref{fig:realparts} shows both quantities as a function of the absolute parameter values.
We note the significant shift of the oscillation frequency relative to the natural membrane frequency ($\omega = 1$) that occurs for some parameter combinations.

\subsection{Finite amplitude ansatz}

Close to the Hopf bifurcation, for small amplitudes, the frequency of the self-sustained oscillations follow from the local analysis of the equations of motion via the stability matrix just presented.
We now develop an analytical description to understand the properties of the self-sustained oscillations also at finite amplitudes, away from the Hopf bifurcation.

The starting point is the ansatz
\begin{equation}\label{XAnsatz}
 x(t) =  x_c + A \cos (\omega t + \vartheta)
\end{equation}
for a simple periodic membrane oscillation at amplitude $A$.
In contrast to the ansatz for the ``cantilever-cavity'' system with one photon mode~\cite{MHG06,LKM08},
where self-sustained oscillations occur at the natural cantilever frequency,
the oscillation frequency $\omega$ has to be included as a parameter in our ansatz,
because in general $\omega \ne 1$ already in the vicinity of the Hopf bifurcation.
The phase angle $\vartheta$ is arbitrary and set to $\vartheta=0$.

With the periodic ansatz~\eqref{XAnsatz} for the membrane position also the optical modes 
follow a periodic motion,
 but additional sidebands at multiples of $\omega$ occur.
From the equations of motion~\eqref{EOM3},~\eqref{EOM4} we obtain the Fourier series
\begin{equation}\label{FourierA}
\begin{split}
 a_L(t) &= \phantom{-}e^{-\ii (A/\omega) \sin \omega t} \sum_{n=-\infty}^\infty a_L^n e^{\ii n \omega t}  \;,\\
 a_R(t) &= e^{+\ii (A/\omega) \sin \omega t} \sum_{n=-\infty}^\infty a_R^n e^{\ii n \omega t} \;,
\end{split}
\end{equation}
where the Fourier coefficients fulfill
\begin{subequations}\label{FourierACoeff}
\begin{align}
 a_L^n = \frac{\hat J_n(\frac{A}{\omega}) + J \sum_{m \ne 0}   \hat J_{n-m}(2 \frac{A}{\omega}) a_R^m}{ \Delta - x_c - n \omega + \ii \kappa}  \;,\\[0.5ex]
 a_R^n = \frac{\hat J_n(-\frac{A}{\omega}) + J \sum_{m \ne 0}  \hat J_{n-m}(-2 \frac{A}{\omega}) a_L^m}{ \Delta + x_c - n \omega + \ii \kappa}
\end{align}
\end{subequations}
(see App.~\ref{app:Fourier} for the derivation).
For $J=0$ both equations decouple and directly give the Fourier coefficients in terms of the Bessel functions $\hat J_n(\cdot)$,
but for $J \ne 0$ a coupled system of linear equations has to be solved.
For small $|J|$ this can be done iteratively.

To determine the parameters $x_c$, $A$, $\omega$ in the ansatz we have to insert Eqs.~\eqref{XAnsatz},~\eqref{FourierA} into the first two equations of motion~\eqref{EOM1},~\eqref{EOM2},
which gives the conditions
\begin{subequations}%
\label{CondAnsatz}
\begin{align}
 x_c &=  -g \sum_m |a_L^m|^2 - |a_R^m|^2  \;,  \label{CondX0} \\
\Gamma \omega A &= -2 g \Im \sum_m {a_L^m}^* a_{L}^{m-1} -  {a_R^m}^* a_{R}^{m-1}  \;, \label{CondPower} \\
  A (1-\omega^2) &= - 2 g \Re\sum_m {a_L^m}^* a_{L}^{m-1} -  {a_R^m}^* a_{R}^{m-1}  \;.  \label{CondPhase}
\end{align}
\end{subequations}
The first condition follows from comparison of the Fourier mode $n=0$ on both sides of the equations,
the other two conditions for the Fourier modes $n= \pm 1$.
The contribution of the higher Fourier modes to the membrane motion is neglible within the limits of validity of the ansatz, and they do not give additional conditions.
 
For an intuitive physical interpretation of the three conditions~\eqref{CondAnsatz} we note that Eqs.~\eqref{EOM1},~\eqref{EOM2} are equations of motion of a driven harmonic oscillator, where the driving force is the radiation pressure ($\propto g$).
In this picture, Eq.~\eqref{CondX0} is a condition on the vanishing of the net force acting on the oscillator over one oscillation period, which can be written as $0 = \int_t^{t+2\pi/\omega} \dot p(t') \mathrm dt'$.
Condition~\eqref{CondPower} is a condition on the vanishing of the net change of the 
oscillator energy $E = (x^2 + p^2)/2$ over one oscillation period,
which can be written as $0 =  \delta E =  \int_t^{t+2\pi/\omega} x(t') \dot x(t') + p(t') \dot p(t') \mathrm dt'$.
This allows us to interpret Eq.~\eqref{CondPower} as the power balance
\begin{equation}
 \mathcal P = \mathcal P_\mathrm{rad} - \mathcal P_\mathrm{fric}
\end{equation}
between the energy gain from the radiation pressure acting on the membrane $\mathcal{P}_\mathrm{rad} = -g \omega A \Im \sum_m {a_L^m}^* a_{L}^{m-1} -  {a_R^m}^* a_{R}^{m-1} $ and the average energy loss due to friction $\mathcal{P}_\mathrm{fric} = \Gamma \omega^2A^2/2$.
These first two conditions are equivalent to those introduced in Ref.~\cite{MHG06} for the optomechanical system with one photon mode.

The third new condition~\eqref{CondPhase} can be interpreted as a condition on the net phase shift per oscillation period, i.e., as the condition that $\vartheta$ is constant in Eq.~\eqref{XAnsatz}.
It can be written as $0 = \int_t^{t+2\pi/\omega} (x(t') - x_c) \dot p(t') + \dot x(t') p(t')  \mathrm dt'$.
This condition allows us to determine the oscillation frequency $\omega$ in the ansatz.
It would be missing 
if we 
considered a simpler ansatz with fixed $\omega=1$.

\begin{figure}
\hspace*{\fill}
\includegraphics[scale=0.091]{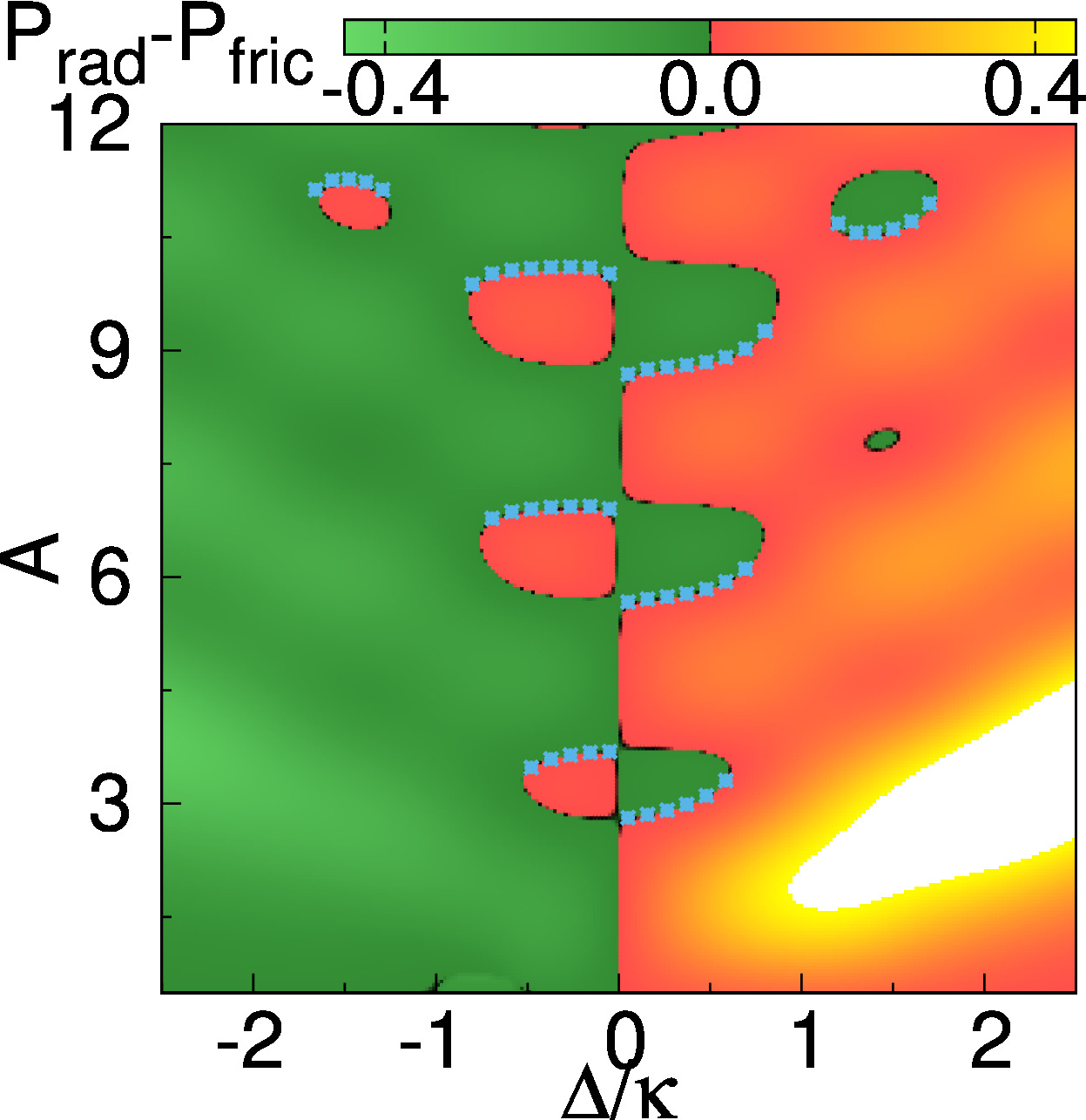}
\hspace*{\fill}
\includegraphics[scale=0.091]{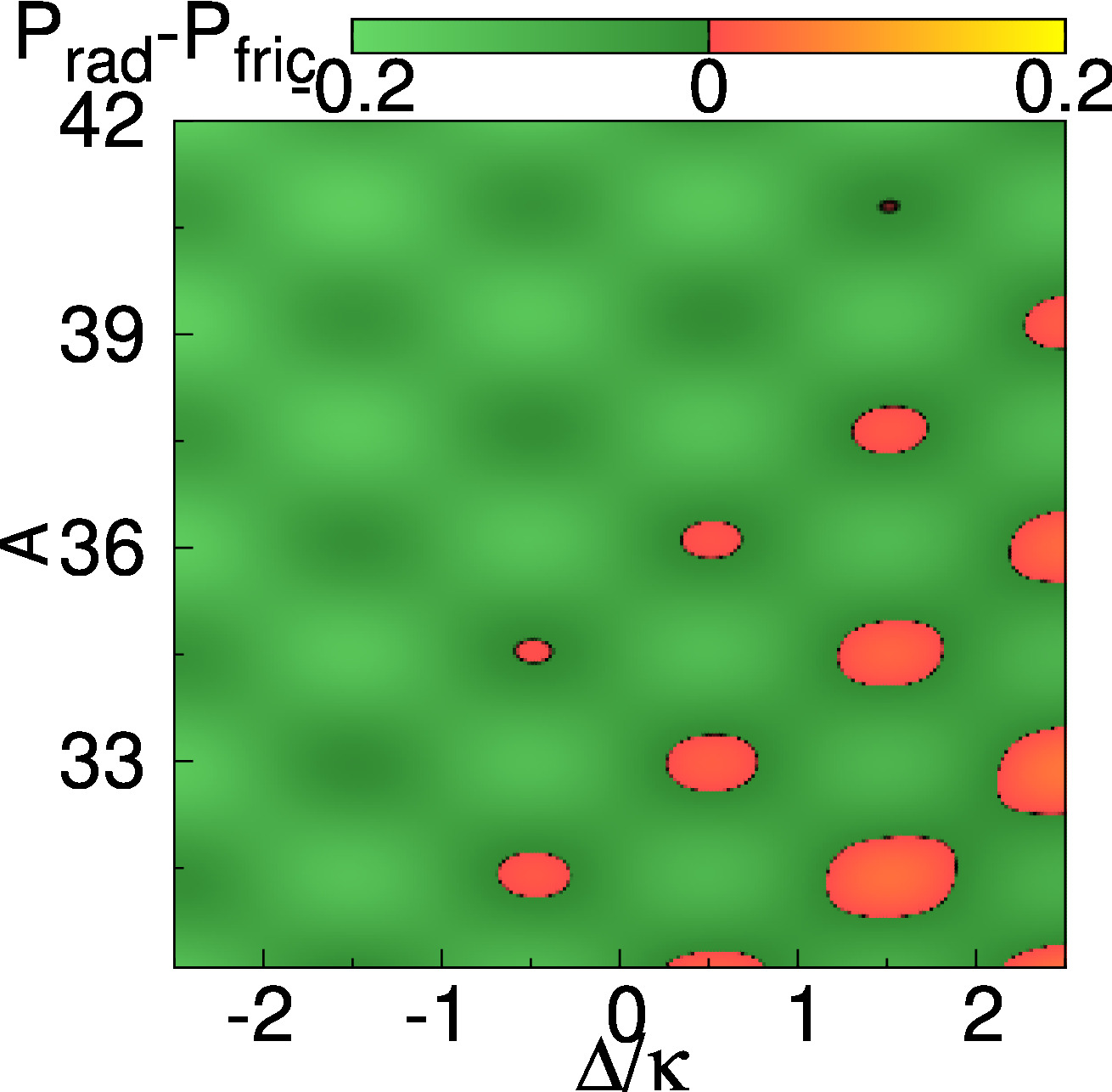}
\hspace*{\fill} \\

\hspace*{\fill}
\includegraphics[scale=0.091]{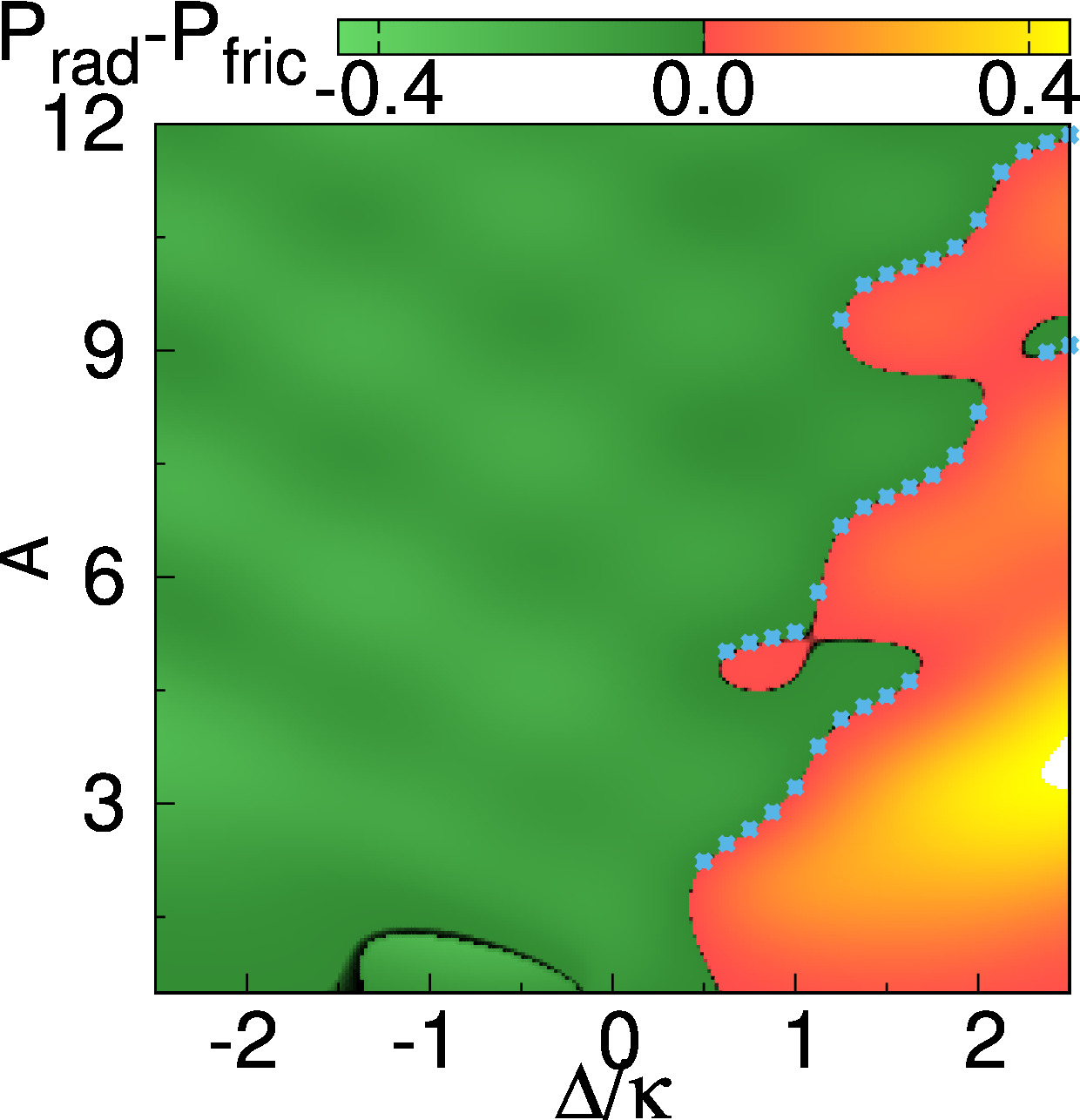}
\hspace*{\fill}
\includegraphics[scale=0.083]{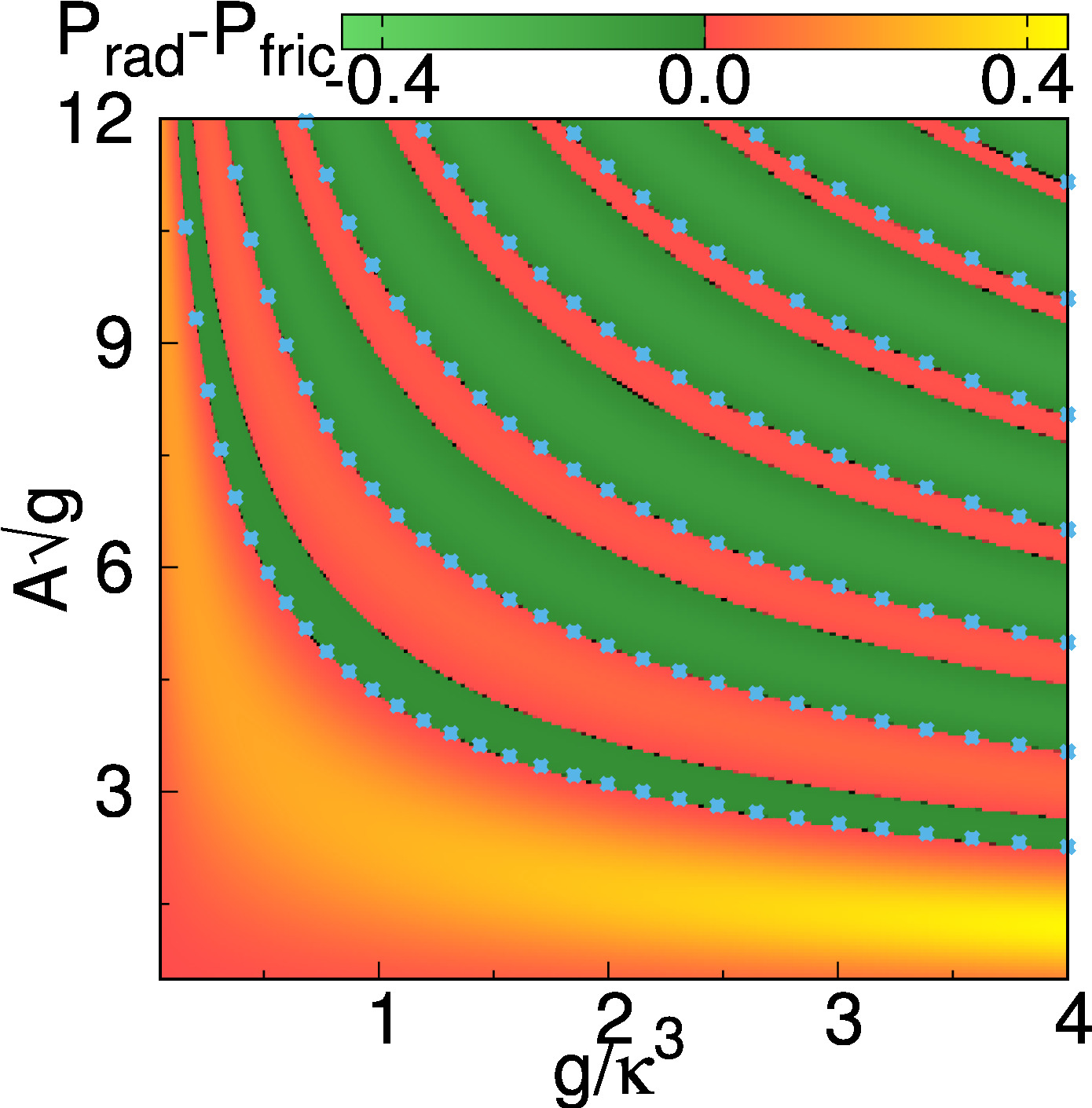}
\hspace*{\fill}\\[-0.2cm]
\caption{(Color online) Power balance $\mathcal P \sim \mathcal{P}_{\text{rad}}-\mathcal{P}_{\text{fric}}$ as a function of the oscillation amplitude $A$ and  $\Delta/\kappa$ ($g/\kappa^3=1$) or $g/\kappa^3$ ($\Delta/\kappa=1.4$), respectively, for $J/\kappa=0$ (upper plots) and $J/\kappa=-0.5$ (lower plots) with $\kappa=1$. Stable periodic orbits obtained from the numerical solution of Eq.~\eqref{EOM} are included as blue dots.
}
\label{fig:PowerBalance}
\end{figure}

The power balance $\mathcal P$ is plotted in Fig.~\ref{fig:PowerBalance}.
For these plots, the oscillation shift $x_c$ and frequency $\omega$ have been determined from the conditions~\eqref{CondX0},~\eqref{CondPhase},
and then the power balance $\mathcal P$ is computed as a function of the remaining free parameter in the ansatz, the oscillation amplitude $A$.
Periodic solutions exist if condition~\eqref{CondPower} is fulfilled, i.e., for $\mathcal P = 0$.
Stable orbits exist if the frictional losses increase with the amplitude, i.e., for $\mathrm d \mathcal P/ \mathrm d A < 0$.

\begin{figure}
\hspace*{\fill}
\includegraphics[scale=0.22]{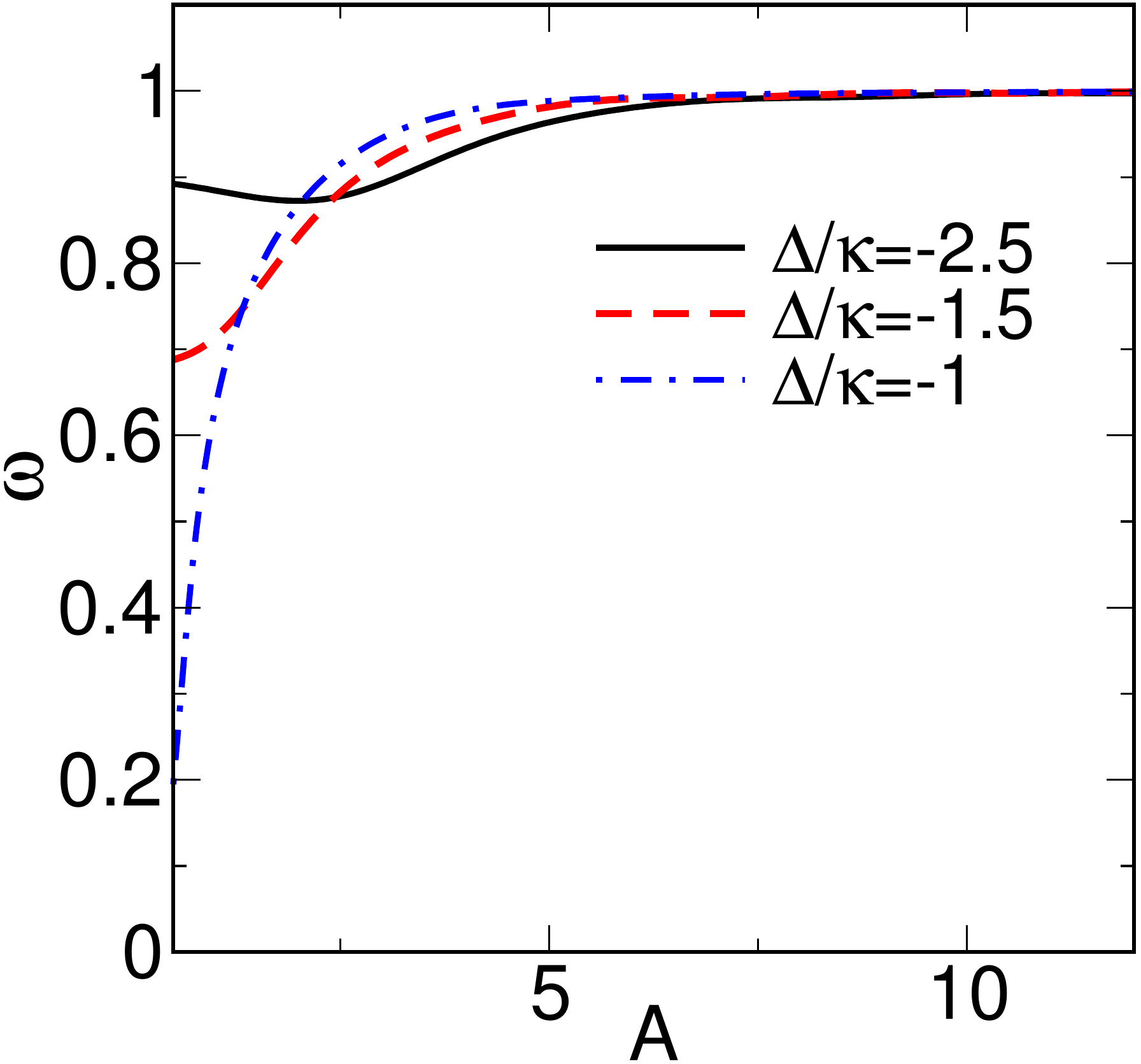}
\hspace*{\fill}
\includegraphics[scale=0.22]{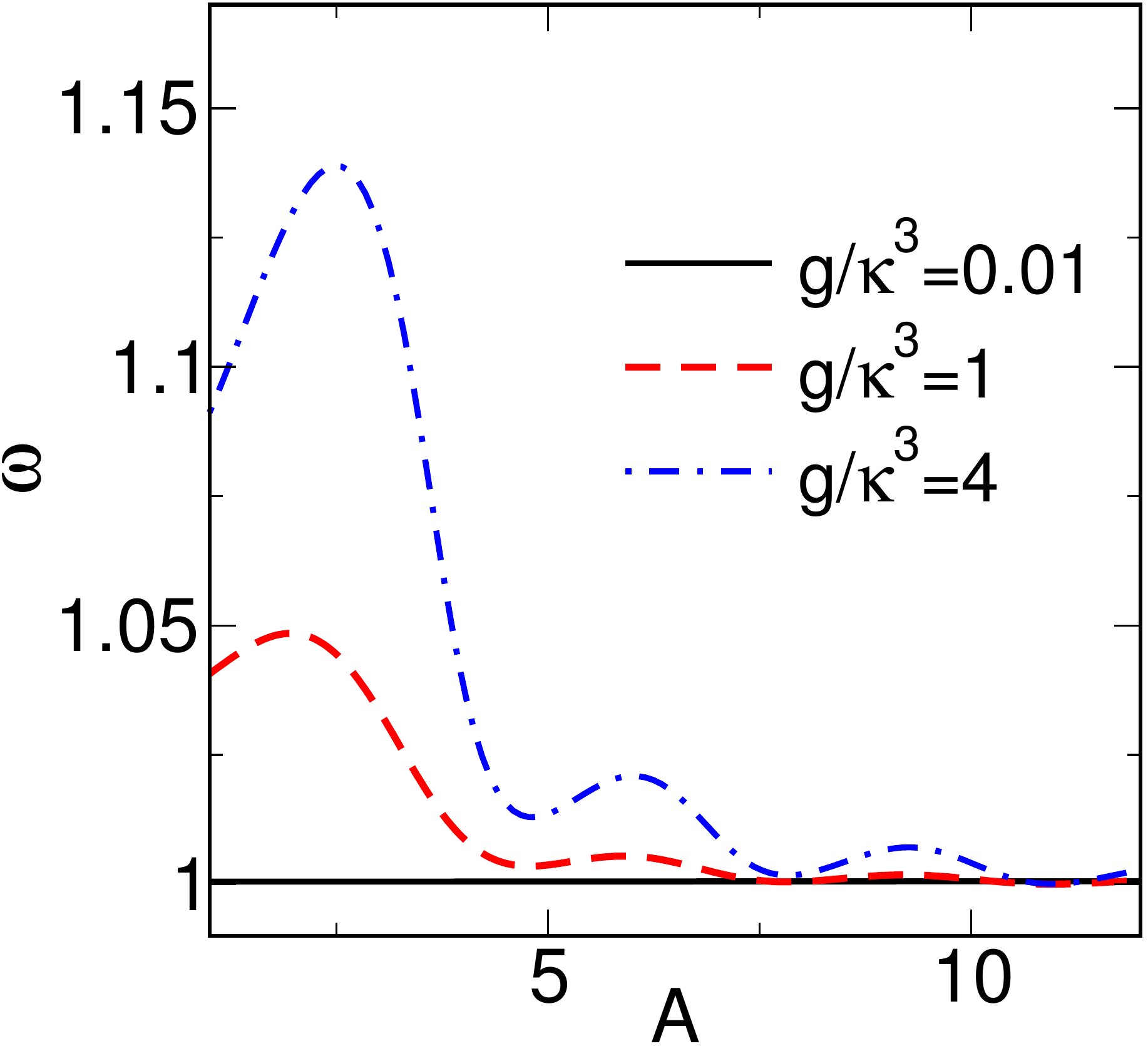}
\hspace*{\fill} \\[2ex]
\hspace*{\fill}
\includegraphics[scale=0.22]{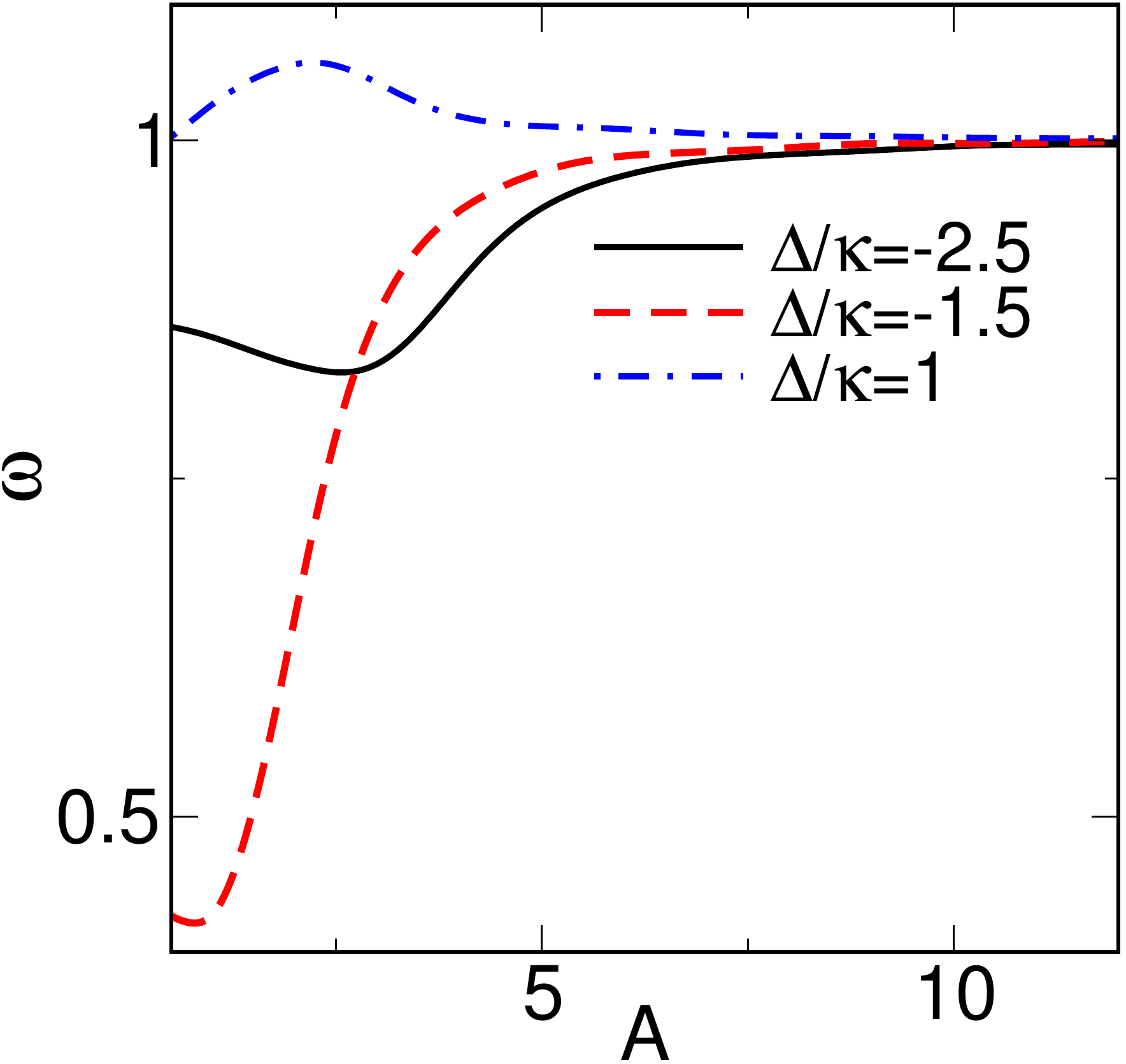}
\hspace*{\fill}
\includegraphics[scale=0.22]{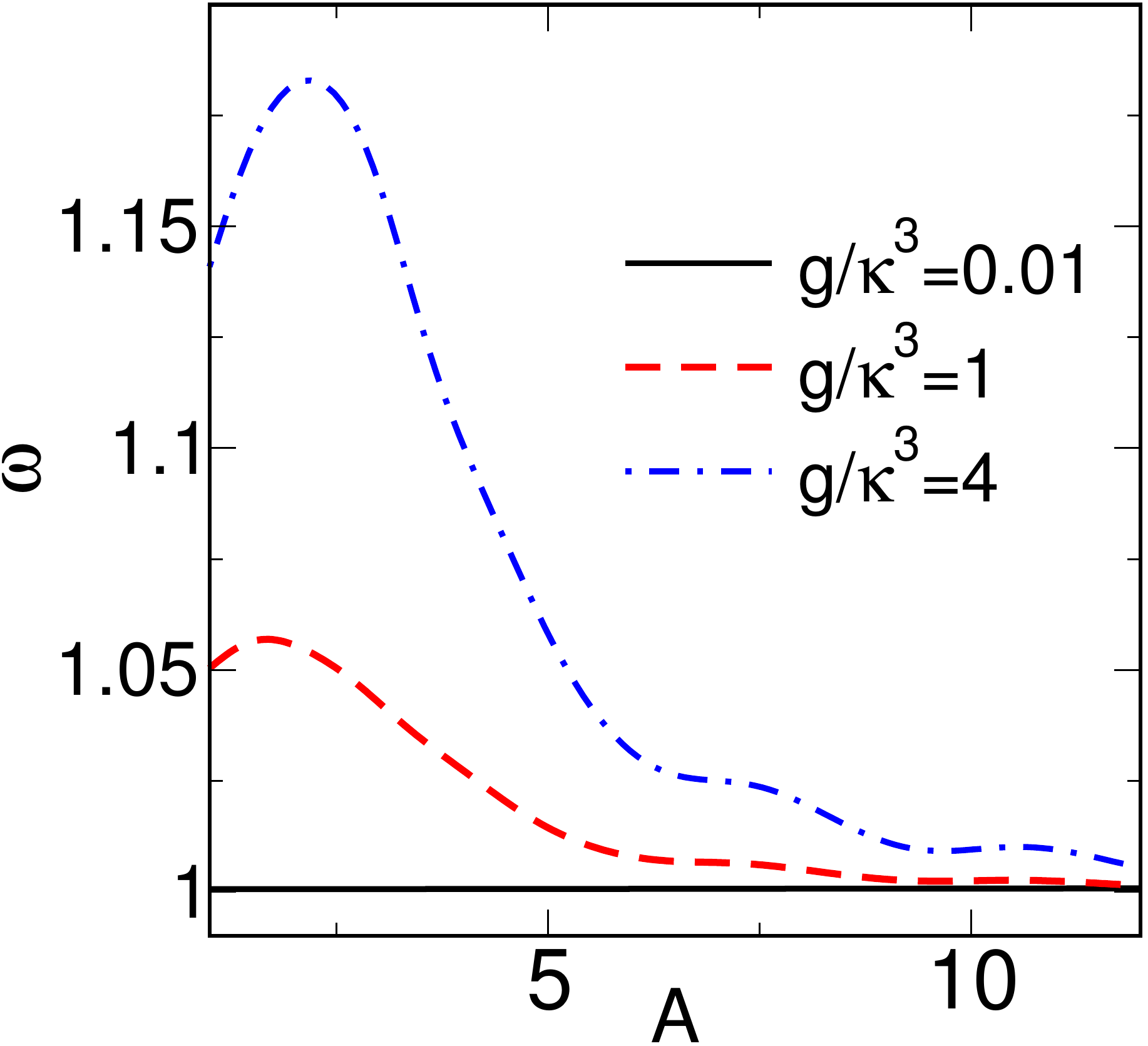}
\hspace*{\fill}\\[-0.15cm]
\caption{(Color online) Oscillation frequency $\omega$  calculated from Eqs.~\eqref{CondAnsatz} as a function of the oscillation amplitude $A$ for $g/\kappa^3=1$ (top/bottom left), $\Delta/\kappa=0.5$ (top right) and $\Delta/\kappa=1.4$ (bottom right) with $\kappa=1$, corresponding to the previous figure (upper plots: $J/\kappa=0$, lower plots: $J/\kappa=-0.5$).
}
\label{fig:OscillationFrequency}
\end{figure}

\begin{figure}
\hspace*{\fill}
\includegraphics[width=0.85\linewidth]{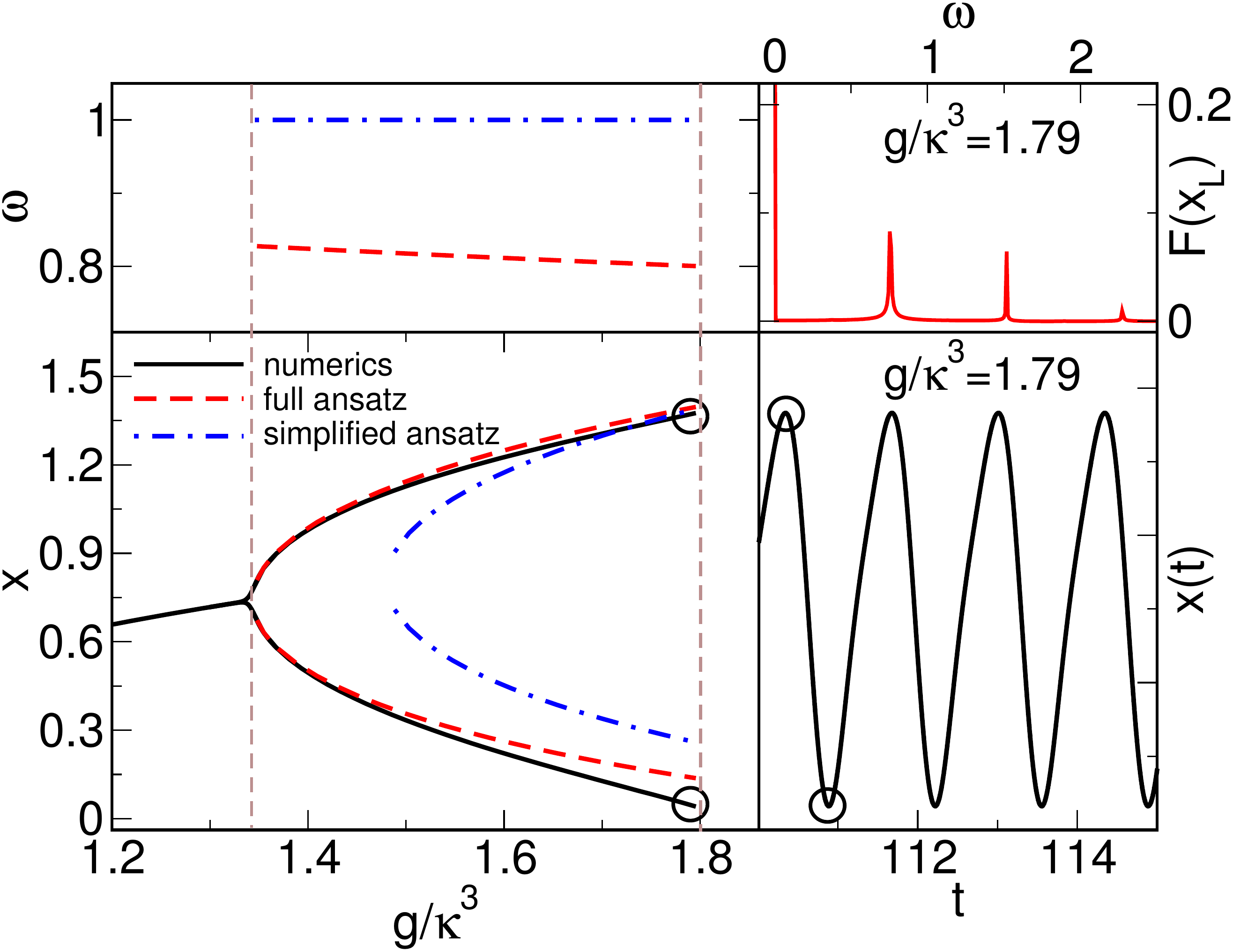}
\hspace*{\fill}
\caption{(Color online) Left panel: 
Maximal and minimal oscillation amplitude $x = x_c \pm A$ obtained with the ansatz~\eqref{XAnsatz} from Eqs.~\eqref{CondAnsatz} (dashed (red) curve) in comparison to values obtained from direct solution of the equations of motion~\eqref{EOM} (solid (black) curve),
for $J/\kappa=0$, $\Delta/\kappa=-0.5$ and $\kappa=1$. 
The upper panel shows the deviation of the oscillation frequency from the bare membrane frequency ($\omega \ne 1$).
Also included are the wrong results obtained with a simplified ansatz with fixed $\omega=1$ (dot-dashed (blue) curves).
Right panel: Cantilever position $x\bk{t}$ (bottom) and the optical spectrum of the left photon mode (top) for $g=1.79$, corresponding to the solution in the left panels (circles). 
\label{fig:AnsatzFail}}
\end{figure}

\paragraph*{Multistability}

For each set of system parameters, i.e., moving parallel to the vertical axis in Fig.~\ref{fig:PowerBalance}, multiple solutions with $P=0$ can be found from the ansatz. Among these, the solutions with
$\mathrm d \mathcal P/ \mathrm d A < 0$ correspond to stable orbits obtained from numerical solution of the equations of motion~\eqref{EOM} (blue dots in Fig.~\ref{fig:PowerBalance}).
Our ansatz thus correctly predicts the the coexistence of multiple stable periodic orbits at different amplitudes,
i.e., the multistability of self-sustained oscillations in the ``membrane-in-the-middle'' setup. 

\paragraph*{Frequency renormalization}
For most parameter combinations the oscillation frequency is shifted significantly relative to the natural membrane frequency (see Fig.~\ref{fig:OscillationFrequency}),
as we noted previously during the analysis of the Hopf bifurcations.
Allowing for $\omega \ne 1$ is crucial to obtain the correct solutions from the ansatz,
while a simpler ansatz with fixed $\omega=1$ would fail (see Fig.~\ref{fig:AnsatzFail}). Since the oscillation frequency $\omega$ appears in Eq.~\eqref{FourierA} for the optical modes it can be observed directly in the optical spectrum  (see Fig.~\ref{fig:AnsatzFail}),
which allows for an experimental measurement.

\section{Route to chaos}
\label{sec:Chaos}

\begin{figure}
\center
\includegraphics[scale=0.29]{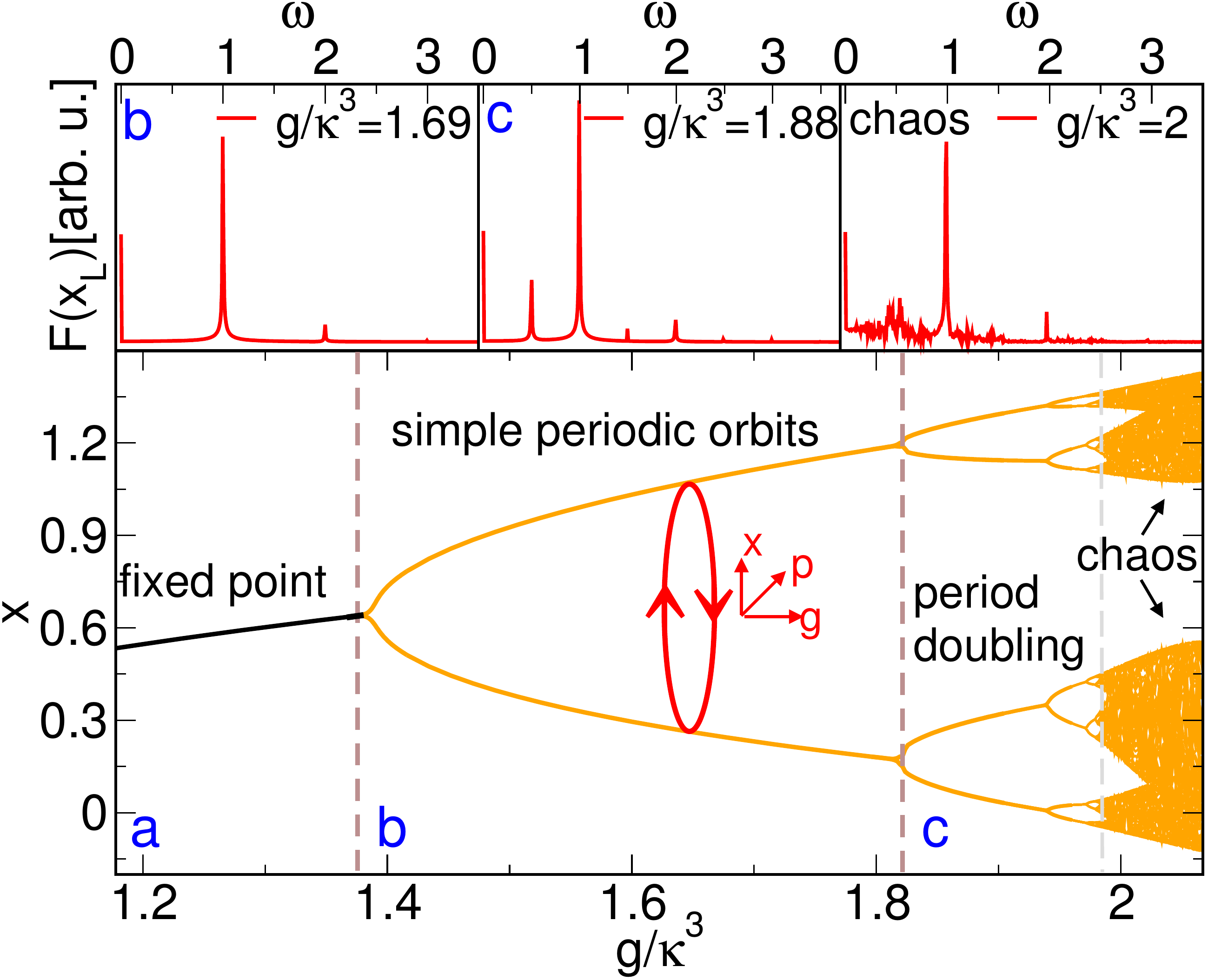}
\caption{(Color online) Feigenbaum diagram starting at the upper fixed point after the supercritical pitchfork bifurcation for $\Delta/\kappa=0$, $J/\kappa=-0.5$ and $\kappa=1$. Proceeding from fixed points (regime a) via simple oscillations (regime b) and period doublings (regime c) finally results in chaos. The different dynamical regimes can be distinguished in the optical spectrum (upper panels, for the left cavity mode).
}
\label{fig:Feigenbaum}
\end{figure}

\begin{figure}
\hspace*{\fill}
\includegraphics[width=0.47\linewidth]{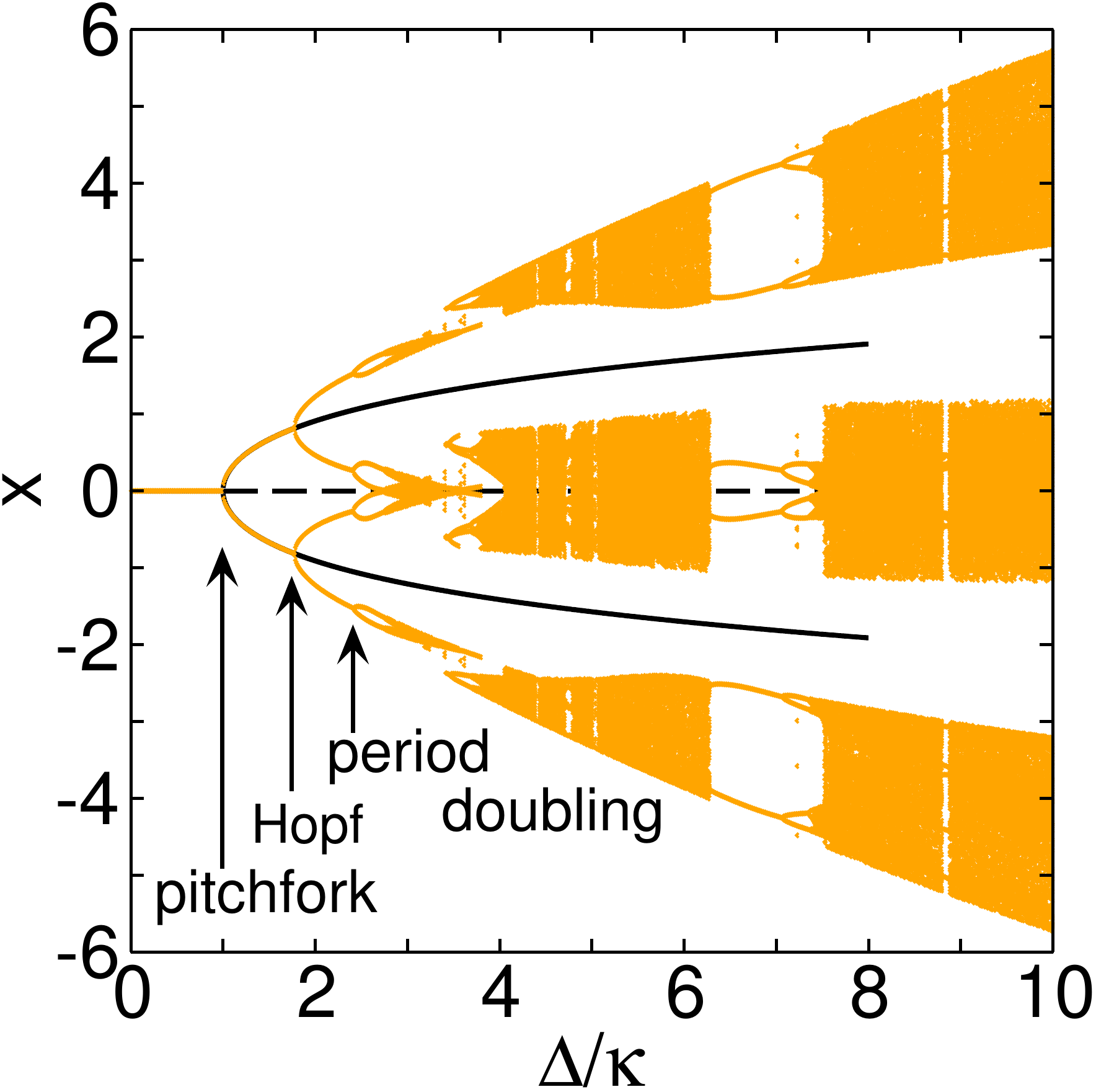}
\hspace*{\fill}
\includegraphics[width=0.47\linewidth]{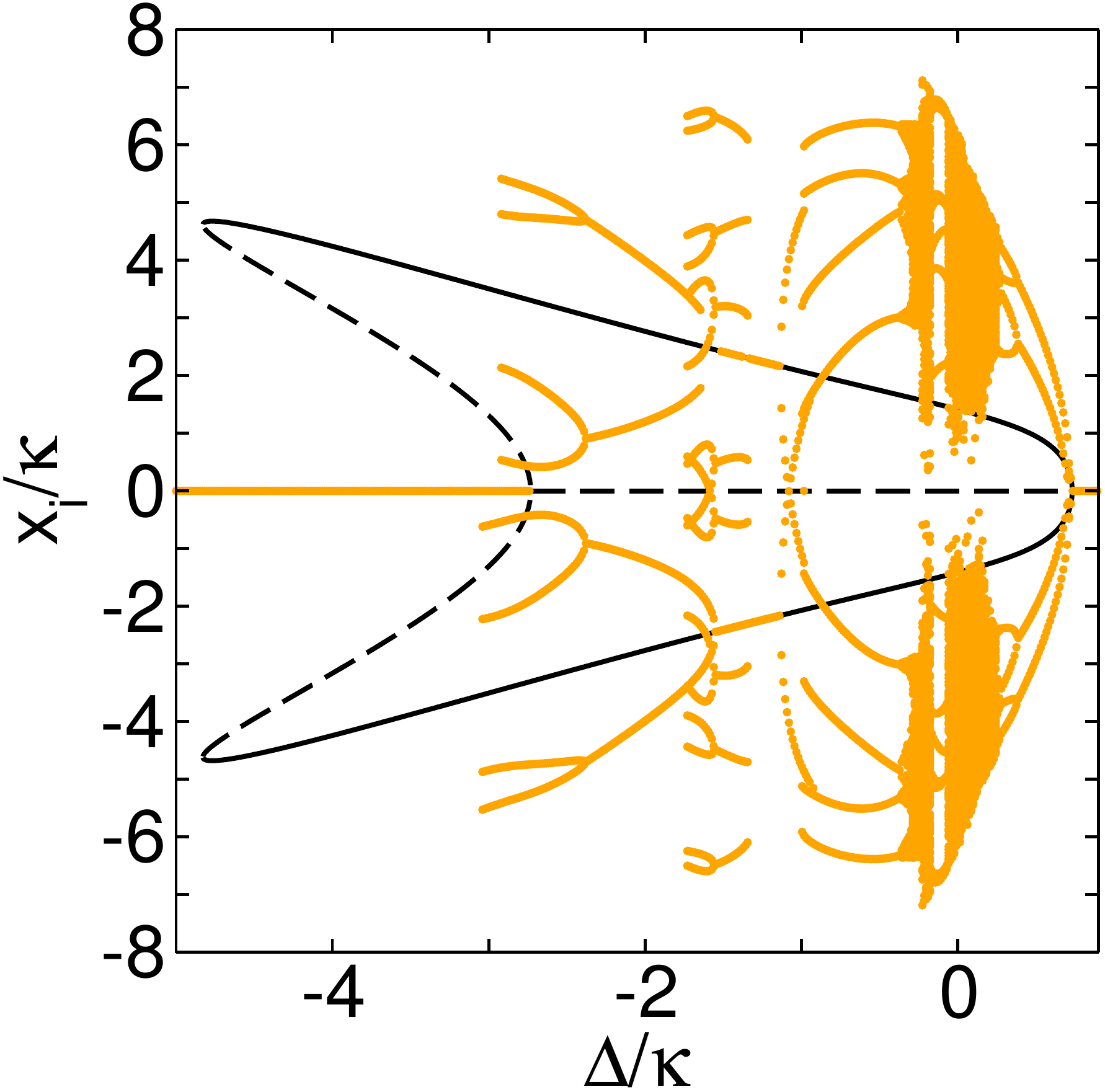}
\hspace*{\fill}
\caption{(Color online) Left panel: Evolution of chaos for small detuning, for $\Delta/\kappa = 0$, $J/\kappa=-1$, $\kappa=1$ as in Fig.~\ref{fig:realparts}.
Right panel: Chaos in the boomerang, for $g/\kappa^3=4$, $J/\kappa=-1$ as in Fig.~\ref{fig:boomerang}, and $\kappa=0.2$.}
\label{fig:ChaosBoomerang}
\end{figure}

Starting from the self-sustained oscillations 
the entire route to chaos in optomechanical systems~\cite{BAF14_PRL} can be observed also for the ``membrane-in-the-middle'' setup.
Fig.~\ref{fig:Feigenbaum} shows the ``Feigenbaum cascade'' of period doubling bifurcations that lead to chaos, starting from the nontrivial fixed point $x_1$ after the supercritical pitchfork bifurcation (cf. Fig.~\ref{fig:bifurcation}).
The sequence of period doubling bifurcations can be observed through the appearance of additional sidebands in the optical spectrum (see upper panels in Fig.~\ref{fig:Feigenbaum}).
Intricate patterns of intertwined regular and chaotic motion replace the fixed point patterns at a larger scale, as shown in Fig.~\ref{fig:ChaosBoomerang} for the supercritical pitchfork bifurcation and the ``boomerang'' pattern. It will probably be hard to resolve details of these features in the experiment,
but it should be possible to measure the position of the first few bifurcations accurately.

\section{Conclusions}
\label{sec:Conc}

A membrane inside a cavity with reflection symmetry shows a variety of fixed point bifurcations related to symmetry breaking.
In addition to symmetry breaking self-sustained oscillations appear for sufficient laser power.
We here analyse the Hopf bifurcations that lead to their existence, and describe their properties with a physically motivated ansatz for finite amplitude oscillations.
The ansatz extends the results obtained for the ``cantilever-cavity'' system with one optical mode~\cite{MHG06,LKM08}
to the situation of two coupled optical modes,
and to oscillations with variable frequency.
The ansatz equations allow for an intuitive interpretation in physical terms,
and especially the power balance proves useful for the prediction of the oscillation amplitudes and frequencies.

In contrast to the ``cantilever-cavity'' system the frequency of the self-sustained oscillations observed here differs from the natural mechanical (i.e., membrane) frequency. 
An interesting promise for future experiments is the indirect measurement of mechanical system parameters, e.g., the membrane stiffness, from the sidebands in the optical spectrum whose position is determined by the frequency shift.
However, a major challenge for the experimental realization of the situation considered in this paper is to achieve the regime of high membrane reflectivity, i.e., small $J$.

In the present paper we focus specifically on the classical dynamics resulting from symmetry-breaking bifurcations.
Certainly, the results reported here are only part of the broader picture of the dynamics of the ``membrane-in-the-middle'' system.
The principal theoretical contribution of this work, our ansatz for the self-sustained oscillations, can be adapted to larger values of $J$, where the full dispersion of the cavity modes has to be taken into account, and also to situations without symmetry breaking, where the membrane is not placed in the cavity center.
Based on a modified ansatz the present analysis extends to these scenarios,
where dynamical patterns similar to those discussed here can be observed,
and which may be more easily realized in the experiment.
These extensions should be addressed in a future study.
A more speculative line of thought is to ask for the influence of quantum effects, such as the breaking of symmetry due to quantum fluctuations and noise,
and the ensuing modifications of the classical bifurcations.

\acknowledgments

The authors are grateful to B. Bruhn and L. Bakemeier for discussions leading to the research reported in this work.
This work was financed by Deutsche Forschungsgemeinschaft through SFB 652 (project B5).

\appendix

\section{Derivation of the dimensionless equations of motion}
\label{app:EOM}

We here summarize the relation between the standard equations of motion for the ``membrane-in-the-middle'' setup (cf. Fig.~\ref{fig:Setup}) given, e.g., in Ref.~\cite{JS08}, and our dimensionless Eqs.~\eqref{EOM}.
Note that we require only the classical equations of motion.
The corresponding Hamiltonian can be constructed according to, e.g., Refs.~\cite{CL11,Law95}.

The equations of motion for the photon amplitudes in the left ($a_l$) and right ($a_r$) half of the cavity, in a reference frame rotating with the laser frequency, have the form
\begin{subequations}%
\begin{align}
\dot{a}_{\text{L}} & =  \Bk{\ii\Delta - \ii G x -\kappa} a_{\text{L}}- \ii Ja_{\text{R}} - \phantom{\sigma} \ii \alpha \;, \\
\dot{a}_{\text{R}} & = \Bk{\ii\Delta + \ii G x -\kappa} a_{\text{R}}- \ii Ja_{\text{L}} -  \sigma \ii \alpha \;,
\end{align}
\end{subequations}
where $\Delta=\Omega_{\text{las}}-\Omega_{\text{cav}}$ is the detuning between the laser frequency $\Omega_{\text{las}}$ and the cavity frequency $\Omega_{\text{cav}} =  n (2\pi c)/(L/2)$ (for the $n$-th optical mode),
and $\kappa$ the cavity decay rate.
In the units chosen here, $|a_{L/R}|^2$ is the number of photons, and $\hbar \Omega_\text{cav} |a_{L/R}|^2$ the energy per optical mode.
$G = - \partial \Omega_{\text{cav}} / \partial x = \Omega_{\text{cav}} / ( L/2 )$ 
gives the change of the optical frequency with membrane position $x$ in the linear regime of small $x$, 
which also determines the radiation pressure.

The parameter $J$, the membrane transmissivity, can be determined from comparison of the position of the optical resonances at $\pm \sqrt{J^2 + G^2 x^2}$ (for $\kappa =0$)
to the quadratic dispersion near $x=0$ obtained from Maxwell's equations~\cite{JS08}.
Note that this treatment is valid only in the limit of small $J$.

The parameter $\alpha$ is related to the laser input power $P$ transmitted into the cavity.
Especially at resonance
we have $|\alpha| = (\kappa P / (2 \hbar \Omega_\mathrm{cav})  )^{1/2}$,
such that the energy per optical mode is $\hbar \Omega_\mathrm{cav} |\alpha/\kappa|^2 =  P/(2 \kappa)$,
in accordance with the choice of $\kappa$ as the amplitude decay rate.
The phase difference $\varphi$ between the laser in each half of the cavity is included through the factor $\sigma=\euler{i\varphi}$.
In the present symmetric setup we consider only phase differences $\varphi=\left\{0,\pi \right\}$,
such that $\sigma=\pm 1$.

The equation of motion for the membrane position ($x$) has the form
\begin{equation}
  \ddot x(t) = - \Omega^2 x(t) - \Gamma \dot x(t) - \hbar (G/m) (|a_L|^2 - |a_R|^2) \;,
\end{equation}
with the membrane frequency $\Omega$, mass $m$, mechanical damping $\Gamma$,
and the radiation pressure $\propto G$.

 To obtain the dimensionless Eqs.~\eqref{EOM}, we now set $\bar x = (G/\Omega) x$,
$\bar p = (G/\Omega^2) \dot x$,
$\bar a_{\mathrm L} = (\Omega/\alpha) a_{\mathrm L}$,
$\bar a_{\mathrm R} = \sigma (\Omega/\alpha) a_{\mathrm R}$,
measure time as $\bar t = \Omega t$,
and define the dimensionless parameters
$\bar \Gamma = \Gamma/\Omega$, $\bar \kappa = \kappa/\Omega$,
$\bar J = \sigma J/\Omega$, $\bar \Delta = \Delta / \Omega$,
$\bar g = 2 \Omega_\text{cav} \kappa P /  (m \Omega^5 L^2)$.
The relation between the dimensionless model parameters and the physical setup parameters is summarized after Eq.~\eqref{EOM}.
Note that $\hbar$ cancels in these equations, as it must in the classical case.
To simplify notation, the overline $\bar{\phantom{x}}$ annotation is omitted 
in the main text.

\section{Fixed point stability}
\label{app:Jac}

For the linear stability analysis we rewrite the equations of motion~\eqref{EOM}
in terms of the quadratures $x_{\text{L/R}}=(1/2) (a_{\text{L/R}}+a^*_{\text{L/R}})$, $p_{\text{L/R}}=(\ii/2) (a^*_{\text{L/R}}-a_{\text{L/R}})$ 
(defined without the usual factor $1/\sqrt2$) instead of the complex variables $a_{L/R}$.
We then get the equations of motion
\begin{subequations}%
\begin{align}
\dot{ x} & = \phantom{-} p  \;, \\
\dot{ p} & =  -  x  -  \Gamma p -  g \left(  x_\mathrm{L}^2 +  p_\mathrm{L}^2 -  x_\mathrm{R}^2 -  p_\mathrm{R}^2  \right) \;, \\
\dot x_\mathrm{L} &= - ( \Delta -  x) p_\mathrm{L} - \kappa x_\mathrm{L} +  J p_\mathrm{R} \;, \\
\dot p_\mathrm{L} &= \phantom{-} ( \Delta -  x) x_\mathrm{L} - \kappa p_\mathrm{L} -  J x_\mathrm{R} - 1\;, \\
\dot x_\mathrm{R} &=- ( \Delta +  x ) p_\mathrm{R} - \kappa x_\mathrm{R} +  J p_\mathrm{L} \;, \\
\dot p_\mathrm{R} &=\phantom{-} ( \Delta +  x ) x_\mathrm{R} - \kappa p_\mathrm{R} -  J x_\mathrm{L} - 1\;,
\end{align}
\end{subequations}
for six real variables.
The stability analysis of the fixed points requires the Jacobi matrix of the right hand side of these equations,
which is given by
\begin{equation}
 \begin{pmatrix}
  0 & 1 & 0 & 0 & 0 & 0 \\
  -1 & - \Gamma & -2  g  x_L & -2  g  p_L  & +2  g  x_R & +2  g   p_R \\[0.5ex]
  p_\mathrm{L} & 0 & - \kappa & -  \Delta 
  +  x & 0 &  J \\[0.25ex]
    - x_\mathrm{L} & 0 &  \Delta -  x & -  \kappa & - J & 0 \\[0.5ex]
    -p_\mathrm{R} & 0 &  0 &  J & - \kappa & -  \Delta -  x & \\[0.25ex]
      x_\mathrm{R} & 0 & - J & 0 &  \Delta +  x & - \kappa
 \end{pmatrix} 
\end{equation}

\leavevmode

\noindent
with the respective fixed point values inserted.
For the quadratures, they are
\begin{equation}
\begin{split}
 x_{L/R} &= \frac{(\Delta \pm x + J)(-\Delta^2 + \kappa^2 + x^2 + J^2) - 2 \Delta \kappa^2}{(-\Delta^2 + \kappa^2 + x^2 + J^2)^2 + 4 \Delta^2 \kappa^2} \;, \\
 p_{L/R} &= \frac{\kappa (-\Delta^2 + \kappa^2 + x^2 + J^2) + 2 \Delta \kappa(\Delta \pm x + J)}{(-\Delta^2 + \kappa^2 + x^2 + J^2)^2 + 4 \Delta^2 \kappa^2} \;,
\end{split}
\end{equation}
with the plus (or minus) sign for $x_{L}, p_L$ (or $x_R, p_R$).

\leavevmode

\section{Fourier series solution for the finite amplitude ansatz}
\label{app:Fourier}

To solve the vector-valued linear differential equation
\begin{equation}\label{app:ODE}
 \dot{ \mathbf{x} } (t) = (\mathbf{A} + f(t) \mathbf{B}) \mathbf x (t) + \mathbf c
\end{equation}
we write the solutions as
\begin{equation}
 \mathbf x(t) = e^{g(t) \mathbf B } \, \mathbf y(t) \;,
\end{equation}
where $\dot g(t) = f(t)$.
The vector $\mathbf y(t)$ has to fulfill the differential equation
\begin{equation}
\dot{\mathbf y}(t) = e^{-g(t) \mathbf B} \, \mathbf A \, e^{g(t) \mathbf B} \, \mathbf y(t) \, + \, e^{-g(t) \mathbf B} \mathbf c \;.
\end{equation}
Unless the matrices $\mathbf A$ and $\mathbf B$ commute,
this is a differential equation with time-dependent parameters.

To proceed with the solution, assume that $f(t)$ is a periodic function
without a constant term
such that also $g(t)$ is periodic, say, $g(t+2 \pi/\omega) = g(t)$.
Then, the Fourier expansions
\begin{equation}
\begin{split}
 e^{-g(t) \mathbf B} \, \mathbf A \, e^{g(t) \mathbf B} & = \sum_n e^{\ii n \omega t} \mathbf X_n \;, \\
 e^{-g(t) \mathbf B} \mathbf c &= \sum_n e^{\ii n \omega t} \mathbf c_n
 \end{split}
\end{equation}
give the equations
\begin{equation}
 \mathbf y_n = \frac{1}{\ii \omega n - \mathbf X_0} \big( \mathbf c_n + \sum_{m \ne 0} \mathbf X_m \mathbf y_{n-m} \big) 
\end{equation}
for the Fourier coefficients in the expansion
\begin{equation}
 \mathbf y(t) = \sum_n e^{\ii n \omega t} \mathbf y_n \;.
\end{equation}

Applied to the equations of motion~\eqref{EOM3},~\eqref{EOM4},
with 
\begin{gather}
\mathbf A = \begin{pmatrix} \ii (\Delta - x_c) - \kappa & -\ii J \\ -\ii J & \ii (\Delta + x_c) - \kappa  \end{pmatrix} \;,
\quad \\[0.5ex]
\mathbf B = \begin{pmatrix} -\ii & 0 \\ 0 & \ii \end{pmatrix} \;,
 \quad
\mathbf c = \begin{pmatrix} -\ii \\ -\ii \end{pmatrix} \;,
\end{gather}
and $f(t) = A \cos \omega t$ according to the ansatz~\eqref{XAnsatz},
we have
\begin{equation}
\mathbf X_0 = \begin{pmatrix} \ii (\Delta - x_c) - \kappa & 0 \\ 0 & \ii (\Delta + x_c) - \kappa \end{pmatrix} \;,
\quad 
\end{equation}
\begin{equation}
\mathbf X_m = -\ii J  \begin{pmatrix} 0 &  \hat J_n(2\frac{A}{\omega})  \\  \hat J_n(-2\frac{A}{\omega}) & 0 \end{pmatrix} \;,
\quad 
\end{equation}
\begin{equation}
\mathbf c_n = -\ii \begin{pmatrix}  \hat J_n(\frac{A}{\omega})  \\   \hat J_n(-\frac{A}{\omega}) \end{pmatrix} \;,
\quad 
\end{equation}
with the help of the Jacobi-Anger expansion
\begin{equation}
 e^{\ii z \sin \omega t} = \sum_{n=-\infty}^\infty \hat J_n(z) e^{\ii n \omega t}
\end{equation}
for the Bessel functions $\hat J_n(\cdot)$,
and thus obtain the Fourier coefficients in Eq.~\eqref{FourierACoeff}.


\begin{thebibliography}{30}%
\makeatletter
\providecommand \@ifxundefined [1]{%
 \@ifx{#1\undefined}
}%
\providecommand \@ifnum [1]{%
 \ifnum #1\expandafter \@firstoftwo
 \else \expandafter \@secondoftwo
 \fi
}%
\providecommand \@ifx [1]{%
 \ifx #1\expandafter \@firstoftwo
 \else \expandafter \@secondoftwo
 \fi
}%
\providecommand \natexlab [1]{#1}%
\providecommand \enquote  [1]{``#1''}%
\providecommand \bibnamefont  [1]{#1}%
\providecommand \bibfnamefont [1]{#1}%
\providecommand \citenamefont [1]{#1}%
\providecommand \href@noop [0]{\@secondoftwo}%
\providecommand \href [0]{\begingroup \@sanitize@url \@href}%
\providecommand \@href[1]{\@@startlink{#1}\@@href}%
\providecommand \@@href[1]{\endgroup#1\@@endlink}%
\providecommand \@sanitize@url [0]{\catcode `\\12\catcode `\$12\catcode
  `\&12\catcode `\#12\catcode `\^12\catcode `\_12\catcode `\%12\relax}%
\providecommand \@@startlink[1]{}%
\providecommand \@@endlink[0]{}%
\providecommand \url  [0]{\begingroup\@sanitize@url \@url }%
\providecommand \@url [1]{\endgroup\@href {#1}{\urlprefix }}%
\providecommand \urlprefix  [0]{URL }%
\providecommand \Eprint [0]{\href }%
\providecommand \doibase [0]{http://dx.doi.org/}%
\providecommand \selectlanguage [0]{\@gobble}%
\providecommand \bibinfo  [0]{\@secondoftwo}%
\providecommand \bibfield  [0]{\@secondoftwo}%
\providecommand \translation [1]{[#1]}%
\providecommand \BibitemOpen [0]{}%
\providecommand \bibitemStop [0]{}%
\providecommand \bibitemNoStop [0]{.\EOS\space}%
\providecommand \EOS [0]{\spacefactor3000\relax}%
\providecommand \BibitemShut  [1]{\csname bibitem#1\endcsname}%
\let\auto@bib@innerbib\@empty
\bibitem [{\citenamefont {Kippenberg}\ and\ \citenamefont
  {Vahala}(2008)}]{KV08}%
  \BibitemOpen
  \bibfield  {author} {\bibinfo {author} {\bibfnamefont {T.~J.}\ \bibnamefont
  {Kippenberg}}\ and\ \bibinfo {author} {\bibfnamefont {K.~J.}\ \bibnamefont
  {Vahala}},\ }\href@noop {} {\bibfield  {journal} {\bibinfo  {journal}
  {Science}\ }\textbf {\bibinfo {volume} {321}},\ \bibinfo {pages} {1172}
  (\bibinfo {year} {2008})}\BibitemShut {NoStop}%
\bibitem [{\citenamefont {Marquardt}\ and\ \citenamefont
  {Girvin}(2009)}]{MG09}%
  \BibitemOpen
  \bibfield  {author} {\bibinfo {author} {\bibfnamefont {F.}~\bibnamefont
  {Marquardt}}\ and\ \bibinfo {author} {\bibfnamefont {S.~M.}\ \bibnamefont
  {Girvin}},\ }\href {\doibase 10.1103/Physics.2.40} {\bibfield  {journal}
  {\bibinfo  {journal} {Physics}\ }\textbf {\bibinfo {volume} {2}},\ \bibinfo
  {pages} {40} (\bibinfo {year} {2009})}\BibitemShut {NoStop}%
\bibitem [{\citenamefont {Meystre}(2013)}]{M13}%
  \BibitemOpen
  \bibfield  {author} {\bibinfo {author} {\bibfnamefont {P.}~\bibnamefont
  {Meystre}},\ }\href@noop {} {\bibfield  {journal} {\bibinfo  {journal} {Ann.
  Phys. (Berlin)}\ }\textbf {\bibinfo {volume} {525}},\ \bibinfo {pages} {215}
  (\bibinfo {year} {2013})}\BibitemShut {NoStop}%
\bibitem [{\citenamefont {Aspelmeyer}\ \emph {et~al.}(2014)\citenamefont
  {Aspelmeyer}, \citenamefont {Kippenberg},\ and\ \citenamefont
  {Marquardt}}]{AKM13}%
  \BibitemOpen
  \bibfield  {author} {\bibinfo {author} {\bibfnamefont {M.}~\bibnamefont
  {Aspelmeyer}}, \bibinfo {author} {\bibfnamefont {T.~J.}\ \bibnamefont
  {Kippenberg}}, \ and\ \bibinfo {author} {\bibfnamefont {F.}~\bibnamefont
  {Marquardt}},\ }\href {\doibase 10.1103/RevModPhys.86.1391} {\bibfield
  {journal} {\bibinfo  {journal} {Rev. Mod. Phys.}\ }\textbf {\bibinfo {volume}
  {86}},\ \bibinfo {pages} {1391} (\bibinfo {year} {2014})}\BibitemShut
  {NoStop}%
\bibitem [{\citenamefont {Heinrich}\ \emph {et~al.}(2010)\citenamefont
  {Heinrich}, \citenamefont {Harris},\ and\ \citenamefont {Marquardt}}]{HH10}%
  \BibitemOpen
  \bibfield  {author} {\bibinfo {author} {\bibfnamefont {G.}~\bibnamefont
  {Heinrich}}, \bibinfo {author} {\bibfnamefont {J.~G.~E.}\ \bibnamefont
  {Harris}}, \ and\ \bibinfo {author} {\bibfnamefont {F.}~\bibnamefont
  {Marquardt}},\ }\href {\doibase 10.1103/PhysRevA.81.011801} {\bibfield
  {journal} {\bibinfo  {journal} {Phys. Rev. A}\ }\textbf {\bibinfo {volume}
  {81}},\ \bibinfo {pages} {011801} (\bibinfo {year} {2010})}\BibitemShut
  {NoStop}%
\bibitem [{\citenamefont {Wu}\ \emph {et~al.}(2013)\citenamefont {Wu},
  \citenamefont {Heinrich},\ and\ \citenamefont {Marquardt}}]{WH13}%
  \BibitemOpen
  \bibfield  {author} {\bibinfo {author} {\bibfnamefont {H.}~\bibnamefont
  {Wu}}, \bibinfo {author} {\bibfnamefont {G.}~\bibnamefont {Heinrich}}, \ and\
  \bibinfo {author} {\bibfnamefont {F.}~\bibnamefont {Marquardt}},\ }\href
  {http://stacks.iop.org/1367-2630/15/i=12/a=123022} {\bibfield  {journal}
  {\bibinfo  {journal} {New J. Phys.}\ }\textbf {\bibinfo {volume} {15}},\
  \bibinfo {pages} {123022} (\bibinfo {year} {2013})}\BibitemShut {NoStop}%
\bibitem [{\citenamefont {Rokhsari}\ \emph {et~al.}(2005)\citenamefont
  {Rokhsari}, \citenamefont {Kippenberg}, \citenamefont {Carmon},\ and\
  \citenamefont {Vahala}}]{RKCV05}%
  \BibitemOpen
  \bibfield  {author} {\bibinfo {author} {\bibfnamefont {H.}~\bibnamefont
  {Rokhsari}}, \bibinfo {author} {\bibfnamefont {T.~H.}\ \bibnamefont
  {Kippenberg}}, \bibinfo {author} {\bibfnamefont {T.}~\bibnamefont {Carmon}},
  \ and\ \bibinfo {author} {\bibfnamefont {K.~J.}\ \bibnamefont {Vahala}},\
  }\href@noop {} {\bibfield  {journal} {\bibinfo  {journal} {Opt. Express}\
  }\textbf {\bibinfo {volume} {13}},\ \bibinfo {pages} {5293} (\bibinfo {year}
  {2005})}\BibitemShut {NoStop}%
\bibitem [{\citenamefont {Kippenberg}\ \emph {et~al.}(2005)\citenamefont
  {Kippenberg}, \citenamefont {Rokhsari}, \citenamefont {Carmon}, \citenamefont
  {Scherer},\ and\ \citenamefont {Vahala}}]{KRCSV05}%
  \BibitemOpen
  \bibfield  {author} {\bibinfo {author} {\bibfnamefont {T.~J.}\ \bibnamefont
  {Kippenberg}}, \bibinfo {author} {\bibfnamefont {H.}~\bibnamefont
  {Rokhsari}}, \bibinfo {author} {\bibfnamefont {T.}~\bibnamefont {Carmon}},
  \bibinfo {author} {\bibfnamefont {A.}~\bibnamefont {Scherer}}, \ and\
  \bibinfo {author} {\bibfnamefont {K.~J.}\ \bibnamefont {Vahala}},\ }\href
  {\doibase 10.1103/PhysRevLett.95.033901} {\bibfield  {journal} {\bibinfo
  {journal} {Phys. Rev. Lett.}\ }\textbf {\bibinfo {volume} {95}},\ \bibinfo
  {pages} {033901} (\bibinfo {year} {2005})}\BibitemShut {NoStop}%
\bibitem [{\citenamefont {Carmon}\ \emph {et~al.}(2005)\citenamefont {Carmon},
  \citenamefont {Rokhsari}, \citenamefont {Yang}, \citenamefont {Kippenberg},\
  and\ \citenamefont {Vahala}}]{CRYKV05}%
  \BibitemOpen
  \bibfield  {author} {\bibinfo {author} {\bibfnamefont {T.}~\bibnamefont
  {Carmon}}, \bibinfo {author} {\bibfnamefont {H.}~\bibnamefont {Rokhsari}},
  \bibinfo {author} {\bibfnamefont {L.}~\bibnamefont {Yang}}, \bibinfo {author}
  {\bibfnamefont {T.~J.}\ \bibnamefont {Kippenberg}}, \ and\ \bibinfo {author}
  {\bibfnamefont {K.~J.}\ \bibnamefont {Vahala}},\ }\href {\doibase
  10.1103/PhysRevLett.94.223902} {\bibfield  {journal} {\bibinfo  {journal}
  {Phys. Rev. Lett.}\ }\textbf {\bibinfo {volume} {94}},\ \bibinfo {pages}
  {223902} (\bibinfo {year} {2005})}\BibitemShut {NoStop}%
\bibitem [{\citenamefont {Marquardt}\ \emph {et~al.}(2006)\citenamefont
  {Marquardt}, \citenamefont {Harris},\ and\ \citenamefont {Girvin}}]{MHG06}%
  \BibitemOpen
  \bibfield  {author} {\bibinfo {author} {\bibfnamefont {F.}~\bibnamefont
  {Marquardt}}, \bibinfo {author} {\bibfnamefont {J.~G.~E.}\ \bibnamefont
  {Harris}}, \ and\ \bibinfo {author} {\bibfnamefont {S.~M.}\ \bibnamefont
  {Girvin}},\ }\href {\doibase 10.1103/PhysRevLett.96.103901} {\bibfield
  {journal} {\bibinfo  {journal} {Phys. Rev. Lett.}\ }\textbf {\bibinfo
  {volume} {96}},\ \bibinfo {pages} {103901} (\bibinfo {year}
  {2006})}\BibitemShut {NoStop}%
\bibitem [{\citenamefont {Ludwig}\ \emph {et~al.}(2008)\citenamefont {Ludwig},
  \citenamefont {Kubala},\ and\ \citenamefont {Marquardt}}]{LKM08}%
  \BibitemOpen
  \bibfield  {author} {\bibinfo {author} {\bibfnamefont {M.}~\bibnamefont
  {Ludwig}}, \bibinfo {author} {\bibfnamefont {B.}~\bibnamefont {Kubala}}, \
  and\ \bibinfo {author} {\bibfnamefont {F.}~\bibnamefont {Marquardt}},\ }\href
  {http://stacks.iop.org/1367-2630/10/i=9/a=095013} {\bibfield  {journal}
  {\bibinfo  {journal} {New J. Phys.}\ }\textbf {\bibinfo {volume} {10}},\
  \bibinfo {pages} {095013} (\bibinfo {year} {2008})}\BibitemShut {NoStop}%
\bibitem [{\citenamefont {Carmon}\ \emph {et~al.}(2007)\citenamefont {Carmon},
  \citenamefont {Cross},\ and\ \citenamefont {Vahala}}]{CCV07}%
  \BibitemOpen
  \bibfield  {author} {\bibinfo {author} {\bibfnamefont {T.}~\bibnamefont
  {Carmon}}, \bibinfo {author} {\bibfnamefont {M.~C.}\ \bibnamefont {Cross}}, \
  and\ \bibinfo {author} {\bibfnamefont {K.~J.}\ \bibnamefont {Vahala}},\
  }\href {\doibase 10.1103/PhysRevLett.98.167203} {\bibfield  {journal}
  {\bibinfo  {journal} {Phys. Rev. Lett.}\ }\textbf {\bibinfo {volume} {98}},\
  \bibinfo {pages} {167203} (\bibinfo {year} {2007})}\BibitemShut {NoStop}%
\bibitem [{\citenamefont {Bakemeier}\ \emph {et~al.}(2015)\citenamefont
  {Bakemeier}, \citenamefont {Alvermann},\ and\ \citenamefont
  {Fehske}}]{BAF14_PRL}%
  \BibitemOpen
  \bibfield  {author} {\bibinfo {author} {\bibfnamefont {L.}~\bibnamefont
  {Bakemeier}}, \bibinfo {author} {\bibfnamefont {A.}~\bibnamefont
  {Alvermann}}, \ and\ \bibinfo {author} {\bibfnamefont {H.}~\bibnamefont
  {Fehske}},\ }\href {\doibase 10.1103/PhysRevLett.114.013601} {\bibfield
  {journal} {\bibinfo  {journal} {Phys. Rev. Lett.}\ }\textbf {\bibinfo
  {volume} {114}},\ \bibinfo {pages} {013601} (\bibinfo {year}
  {2015})}\BibitemShut {NoStop}%
\bibitem [{\citenamefont {L\"u}\ \emph {et~al.}(2015)\citenamefont {L\"u},
  \citenamefont {Jing}, \citenamefont {Ma},\ and\ \citenamefont {Wu}}]{LJMW15}%
  \BibitemOpen
  \bibfield  {author} {\bibinfo {author} {\bibfnamefont {X.-Y.}\ \bibnamefont
  {L\"u}}, \bibinfo {author} {\bibfnamefont {H.}~\bibnamefont {Jing}}, \bibinfo
  {author} {\bibfnamefont {J.-Y.}\ \bibnamefont {Ma}}, \ and\ \bibinfo {author}
  {\bibfnamefont {Y.}~\bibnamefont {Wu}},\ }\href {\doibase
  10.1103/PhysRevLett.114.253601} {\bibfield  {journal} {\bibinfo  {journal}
  {Phys. Rev. Lett.}\ }\textbf {\bibinfo {volume} {114}},\ \bibinfo {pages}
  {253601} (\bibinfo {year} {2015})}\BibitemShut {NoStop}%
\bibitem [{\citenamefont {Gutzwiller}(1990)}]{Gutz90}%
  \BibitemOpen
  \bibfield  {author} {\bibinfo {author} {\bibfnamefont {M.~C.}\ \bibnamefont
  {Gutzwiller}},\ }\href@noop {} {\emph {\bibinfo {title} {Chaos in Classical
  and Quantum Mechanics}}}\ (\bibinfo  {publisher} {Springer},\ \bibinfo {year}
  {1990})\BibitemShut {NoStop}%
\bibitem [{\citenamefont {Gardiner}\ and\ \citenamefont {Zoller}(2004)}]{GZ04}%
  \BibitemOpen
  \bibfield  {author} {\bibinfo {author} {\bibfnamefont {C.~W.}\ \bibnamefont
  {Gardiner}}\ and\ \bibinfo {author} {\bibfnamefont {P.}~\bibnamefont
  {Zoller}},\ }\href@noop {} {\emph {\bibinfo {title} {Quantum Noise}}}\
  (\bibinfo  {publisher} {Springer},\ \bibinfo {year} {2004})\BibitemShut
  {NoStop}%
\bibitem [{\citenamefont {Schleich}(2001)}]{Schl01}%
  \BibitemOpen
  \bibfield  {author} {\bibinfo {author} {\bibfnamefont {W.~P.}\ \bibnamefont
  {Schleich}},\ }\href@noop {} {\emph {\bibinfo {title} {Quantum Optics in
  Phase Space}}}\ (\bibinfo  {publisher} {Wiley-VCH},\ \bibinfo {year}
  {2001})\BibitemShut {NoStop}%
\bibitem [{\citenamefont {Armata}\ \emph {et~al.}(2016)\citenamefont {Armata},
  \citenamefont {Latmiral}, \citenamefont {Pikovski}, \citenamefont {Vanner},
  \citenamefont {Brukner},\ and\ \citenamefont {Kim}}]{ALPVBK16}%
  \BibitemOpen
  \bibfield  {author} {\bibinfo {author} {\bibfnamefont {F.}~\bibnamefont
  {Armata}}, \bibinfo {author} {\bibfnamefont {L.}~\bibnamefont {Latmiral}},
  \bibinfo {author} {\bibfnamefont {I.}~\bibnamefont {Pikovski}}, \bibinfo
  {author} {\bibfnamefont {M.~R.}\ \bibnamefont {Vanner}}, \bibinfo {author}
  {\bibfnamefont {C.}~\bibnamefont {Brukner}}, \ and\ \bibinfo {author}
  {\bibfnamefont {M.~S.}\ \bibnamefont {Kim}},\ }\href {\doibase
  10.1103/PhysRevA.93.063862} {\bibfield  {journal} {\bibinfo  {journal} {Phys.
  Rev. A}\ }\textbf {\bibinfo {volume} {93}},\ \bibinfo {pages} {063862}
  (\bibinfo {year} {2016})}\BibitemShut {NoStop}%
\bibitem [{\citenamefont {Wang}\ \emph {et~al.}(2016)\citenamefont {Wang},
  \citenamefont {Lai},\ and\ \citenamefont {Grebogi}}]{WLG16}%
  \BibitemOpen
  \bibfield  {author} {\bibinfo {author} {\bibfnamefont {G.}~\bibnamefont
  {Wang}}, \bibinfo {author} {\bibfnamefont {Y.-C.}\ \bibnamefont {Lai}}, \
  and\ \bibinfo {author} {\bibfnamefont {C.}~\bibnamefont {Grebogi}},\ }\href
  {\doibase 10.1038/srep35381} {\bibfield  {journal} {\bibinfo  {journal} {Sci.
  Rep.}\ }\textbf {\bibinfo {volume} {6}},\ \bibinfo {pages} {35381} (\bibinfo
  {year} {2016})}\BibitemShut {NoStop}%
\bibitem [{\citenamefont {Schulz}\ \emph {et~al.}(2016)\citenamefont {Schulz},
  \citenamefont {Alvermann}, \citenamefont {Bakemeier},\ and\ \citenamefont
  {Fehske}}]{SA16}%
  \BibitemOpen
  \bibfield  {author} {\bibinfo {author} {\bibfnamefont {C.}~\bibnamefont
  {Schulz}}, \bibinfo {author} {\bibfnamefont {A.}~\bibnamefont {Alvermann}},
  \bibinfo {author} {\bibfnamefont {L.}~\bibnamefont {Bakemeier}}, \ and\
  \bibinfo {author} {\bibfnamefont {H.}~\bibnamefont {Fehske}},\ }\href
  {http://stacks.iop.org/0295-5075/113/i=6/a=64002} {\bibfield  {journal}
  {\bibinfo  {journal} {Europhys. Lett.}\ }\textbf {\bibinfo {volume} {113}},\
  \bibinfo {pages} {64002} (\bibinfo {year} {2016})}\BibitemShut {NoStop}%
\bibitem [{\citenamefont {Mumford}\ \emph {et~al.}(2015)\citenamefont
  {Mumford}, \citenamefont {O'Dell},\ and\ \citenamefont {Larson}}]{MO15}%
  \BibitemOpen
  \bibfield  {author} {\bibinfo {author} {\bibfnamefont {J.}~\bibnamefont
  {Mumford}}, \bibinfo {author} {\bibfnamefont {D.~H.~J.}\ \bibnamefont
  {O'Dell}}, \ and\ \bibinfo {author} {\bibfnamefont {J.}~\bibnamefont
  {Larson}},\ }\href {\doibase 10.1002/andp.201400105} {\bibfield  {journal}
  {\bibinfo  {journal} {Ann. Phys.}\ }\textbf {\bibinfo {volume} {527}},\
  \bibinfo {pages} {115} (\bibinfo {year} {2015})}\BibitemShut {NoStop}%
\bibitem [{\citenamefont {Larson}\ and\ \citenamefont {Horsdal}(2011)}]{LH11}%
  \BibitemOpen
  \bibfield  {author} {\bibinfo {author} {\bibfnamefont {J.}~\bibnamefont
  {Larson}}\ and\ \bibinfo {author} {\bibfnamefont {M.}~\bibnamefont
  {Horsdal}},\ }\href {\doibase 10.1103/PhysRevA.84.021804} {\bibfield
  {journal} {\bibinfo  {journal} {Phys. Rev. A}\ }\textbf {\bibinfo {volume}
  {84}},\ \bibinfo {pages} {021804} (\bibinfo {year} {2011})}\BibitemShut
  {NoStop}%
\bibitem [{\citenamefont {Xu}\ \emph {et~al.}(2015)\citenamefont {Xu},
  \citenamefont {Kemiktarak}, \citenamefont {Fan}, \citenamefont {Ragole},
  \citenamefont {Lawall},\ and\ \citenamefont {Taylor}}]{XKFRLT15}%
  \BibitemOpen
  \bibfield  {author} {\bibinfo {author} {\bibfnamefont {H.}~\bibnamefont
  {Xu}}, \bibinfo {author} {\bibfnamefont {U.}~\bibnamefont {Kemiktarak}},
  \bibinfo {author} {\bibfnamefont {J.}~\bibnamefont {Fan}}, \bibinfo {author}
  {\bibfnamefont {S.}~\bibnamefont {Ragole}}, \bibinfo {author} {\bibfnamefont
  {J.}~\bibnamefont {Lawall}}, \ and\ \bibinfo {author} {\bibfnamefont {J.~M.}\
  \bibnamefont {Taylor}},\ }\href@noop {} {\enquote {\bibinfo {title}
  {Observation of optomechanical buckling phase transitions},}\ } (\bibinfo
  {year} {2015}),\ \bibinfo {note} {arXiv:1510.04971}\BibitemShut {NoStop}%
\bibitem [{\citenamefont {Ruiz-Rivas}\ \emph {et~al.}(2016)\citenamefont
  {Ruiz-Rivas}, \citenamefont {Navarrete-Benlloch}, \citenamefont {Patera},
  \citenamefont {Rold\'an},\ and\ \citenamefont {de~Valc\'arcel}}]{RN16}%
  \BibitemOpen
  \bibfield  {author} {\bibinfo {author} {\bibfnamefont {J.}~\bibnamefont
  {Ruiz-Rivas}}, \bibinfo {author} {\bibfnamefont {C.}~\bibnamefont
  {Navarrete-Benlloch}}, \bibinfo {author} {\bibfnamefont {G.}~\bibnamefont
  {Patera}}, \bibinfo {author} {\bibfnamefont {E.}~\bibnamefont {Rold\'an}}, \
  and\ \bibinfo {author} {\bibfnamefont {G.~J.}\ \bibnamefont
  {de~Valc\'arcel}},\ }\href {\doibase 10.1103/PhysRevA.93.033850} {\bibfield
  {journal} {\bibinfo  {journal} {Phys. Rev. A}\ }\textbf {\bibinfo {volume}
  {93}},\ \bibinfo {pages} {033850} (\bibinfo {year} {2016})}\BibitemShut
  {NoStop}%
\bibitem [{\citenamefont {Thompson}\ \emph {et~al.}(2008)\citenamefont
  {Thompson}, \citenamefont {Zwickl}, \citenamefont {Jayich}, \citenamefont
  {Marquardt}, \citenamefont {Girvin},\ and\ \citenamefont {Harris}}]{TZ07}%
  \BibitemOpen
  \bibfield  {author} {\bibinfo {author} {\bibfnamefont {J.}~\bibnamefont
  {Thompson}}, \bibinfo {author} {\bibfnamefont {B.}~\bibnamefont {Zwickl}},
  \bibinfo {author} {\bibfnamefont {A.}~\bibnamefont {Jayich}}, \bibinfo
  {author} {\bibfnamefont {F.}~\bibnamefont {Marquardt}}, \bibinfo {author}
  {\bibfnamefont {S.~M.}\ \bibnamefont {Girvin}}, \ and\ \bibinfo {author}
  {\bibfnamefont {J.~G.~E.}\ \bibnamefont {Harris}},\ }\href@noop {} {\bibfield
   {journal} {\bibinfo  {journal} {Nature}\ }\textbf {\bibinfo {volume}
  {452}},\ \bibinfo {pages} {72} (\bibinfo {year} {2008})}\BibitemShut
  {NoStop}%
\bibitem [{\citenamefont {Jayich}\ \emph {et~al.}(2008)\citenamefont {Jayich},
  \citenamefont {Sankey}, \citenamefont {Zwickl}, \citenamefont {Yang},
  \citenamefont {Thompson}, \citenamefont {Girvin}, \citenamefont {Clerk},
  \citenamefont {Marquardt},\ and\ \citenamefont {Harris}}]{JS08}%
  \BibitemOpen
  \bibfield  {author} {\bibinfo {author} {\bibfnamefont {A.~M.}\ \bibnamefont
  {Jayich}}, \bibinfo {author} {\bibfnamefont {J.~C.}\ \bibnamefont {Sankey}},
  \bibinfo {author} {\bibfnamefont {B.~M.}\ \bibnamefont {Zwickl}}, \bibinfo
  {author} {\bibfnamefont {C.}~\bibnamefont {Yang}}, \bibinfo {author}
  {\bibfnamefont {J.~D.}\ \bibnamefont {Thompson}}, \bibinfo {author}
  {\bibfnamefont {S.~M.}\ \bibnamefont {Girvin}}, \bibinfo {author}
  {\bibfnamefont {A.~A.}\ \bibnamefont {Clerk}}, \bibinfo {author}
  {\bibfnamefont {F.}~\bibnamefont {Marquardt}}, \ and\ \bibinfo {author}
  {\bibfnamefont {J.~G.~E.}\ \bibnamefont {Harris}},\ }\href
  {http://stacks.iop.org/1367-2630/10/i=9/a=095008} {\bibfield  {journal}
  {\bibinfo  {journal} {New J. Phys.}\ }\textbf {\bibinfo {volume} {10}},\
  \bibinfo {pages} {095008} (\bibinfo {year} {2008})}\BibitemShut {NoStop}%
\bibitem [{\citenamefont {Bhattacharya}\ \emph {et~al.}(2008)\citenamefont
  {Bhattacharya}, \citenamefont {Uys},\ and\ \citenamefont {Meystre}}]{BU08}%
  \BibitemOpen
  \bibfield  {author} {\bibinfo {author} {\bibfnamefont {M.}~\bibnamefont
  {Bhattacharya}}, \bibinfo {author} {\bibfnamefont {H.}~\bibnamefont {Uys}}, \
  and\ \bibinfo {author} {\bibfnamefont {P.}~\bibnamefont {Meystre}},\ }\href
  {\doibase 10.1103/PhysRevA.77.033819} {\bibfield  {journal} {\bibinfo
  {journal} {Phys. Rev. A}\ }\textbf {\bibinfo {volume} {77}},\ \bibinfo
  {pages} {033819} (\bibinfo {year} {2008})}\BibitemShut {NoStop}%
\bibitem [{\citenamefont {Ludwig}\ \emph {et~al.}(2012)\citenamefont {Ludwig},
  \citenamefont {Safavi-Naeini}, \citenamefont {Painter},\ and\ \citenamefont
  {Marquardt}}]{LS12}%
  \BibitemOpen
  \bibfield  {author} {\bibinfo {author} {\bibfnamefont {M.}~\bibnamefont
  {Ludwig}}, \bibinfo {author} {\bibfnamefont {A.~H.}\ \bibnamefont
  {Safavi-Naeini}}, \bibinfo {author} {\bibfnamefont {O.}~\bibnamefont
  {Painter}}, \ and\ \bibinfo {author} {\bibfnamefont {F.}~\bibnamefont
  {Marquardt}},\ }\href {\doibase 10.1103/PhysRevLett.109.063601} {\bibfield
  {journal} {\bibinfo  {journal} {Phys. Rev. Lett.}\ }\textbf {\bibinfo
  {volume} {109}},\ \bibinfo {pages} {063601} (\bibinfo {year}
  {2012})}\BibitemShut {NoStop}%
\bibitem [{\citenamefont {Cheung}\ and\ \citenamefont {Law}(2011)}]{CL11}%
  \BibitemOpen
  \bibfield  {author} {\bibinfo {author} {\bibfnamefont {H.~K.}\ \bibnamefont
  {Cheung}}\ and\ \bibinfo {author} {\bibfnamefont {C.~K.}\ \bibnamefont
  {Law}},\ }\href {\doibase 10.1103/PhysRevA.84.023812} {\bibfield  {journal}
  {\bibinfo  {journal} {Phys. Rev. A}\ }\textbf {\bibinfo {volume} {84}},\
  \bibinfo {pages} {023812} (\bibinfo {year} {2011})}\BibitemShut {NoStop}%
\bibitem [{\citenamefont {Law}(1995)}]{Law95}%
  \BibitemOpen
  \bibfield  {author} {\bibinfo {author} {\bibfnamefont {C.~K.}\ \bibnamefont
  {Law}},\ }\href {\doibase 10.1103/PhysRevA.51.2537} {\bibfield  {journal}
  {\bibinfo  {journal} {Phys. Rev. A}\ }\textbf {\bibinfo {volume} {51}},\
  \bibinfo {pages} {2537} (\bibinfo {year} {1995})}\BibitemShut {NoStop}%
\end{thebibliography}
\end{document}